\documentclass[nojss, nofooter, article, table]{jss}

\usepackage[utf8]{inputenc} %
\usepackage{url}            %
\usepackage{booktabs}       %
\usepackage{tabularx}
\usepackage{amsmath}
\usepackage{amsfonts}       %
\usepackage{nicefrac}       %
\usepackage{microtype}      %
\usepackage{algorithm}
\usepackage{algpseudocode}
\setcitestyle{notesep={}}

\usepackage{bm}

\providecommand{\tightlist}{\setlength{\itemsep}{0pt}\setlength{\parskip}{0pt}}

\title{
\liesel{}: A \proglang{Python} Framework for Graph-Based Bayesian Modeling and Customizable MCMC with Support for Generalized Additive Models
}
\Plaintitle{Liesel: A Python Framework for Graph-Based Bayesian Modeling and Customizable MCMC with Support for Generalized Additive Models
}
\Shorttitle{\liesel: Bayesian Modeling, Customizable MCMC, and GAMs in \proglang{Python}}

\author{%
	Hannes Riebl%
	\\Chair of Statistics\\University of Göttingen
  \And
  Johannes Brachem
  \\Chair of Statistics\\University of Göttingen
  \AND
  Thomas Kneib%
  \\Chair of Statistics\\University of Göttingen
  \And
  Gianmarco Callegher
  \\Chair of Statistics\\University of Göttingen
  \And
  Paul F.V.\ Wiemann%
  \\Department of Statistics\\The Ohio State University
}

\Plainauthor{Riebl, Brachem, Kneib, Callegher, Wiemann}

\Abstract{
\pkg{Liesel} is a \proglang{Python} framework for Bayesian model building and posterior computation with dedicated support for generalized additive regression models that is designed to reduce friction in methodological work. The framework consists of three components. The first component, \lieselmodel{}, represents models as directed acyclic graphs and supports interactive model construction, modification, visualization, prediction, and prior or posterior predictive simulation. The second component, \goose{}, provides a modular MCMC framework based on reusable kernels and supports blocked componentwise sampling as well as user-defined Gibbs and Metropolis--Hastings updates, while leveraging \jax{} for automatic differentiation, just-in-time compilation, and hardware acceleration. The third component, \lieselgam{}, supplies high-level building blocks for generalized additive regression models and provides functionality for formula-based model specification, summaries, diagnostics, and effect visualization. Together, these parts make \pkg{Liesel} an effective tool for the rapid development, testing, and application of Bayesian models and MCMC algorithms. \liesel{}'s modular architecture allows users to extend the software to models and inference algorithms beyond generalized additive models and MCMC, offering considerable flexibility for a wide spectrum of statistical research.
}

\Plainkeywords{Python, Bayesian networks, probabilistic graphical models, Markov chain Monte Carlo, semi-parametric regression, structured additive regression, generalized additive models}

\DeclareBoldMathCommand\xvec{x}
\DeclareBoldMathCommand\pvec{p}
\DeclareBoldMathCommand\yvec{y}

\DeclareBoldMathCommand\alphavec{\alpha}
\DeclareBoldMathCommand\betavec{\beta}
\DeclareBoldMathCommand\gammavec{\gamma}
\DeclareBoldMathCommand\etavec{\eta}
\DeclareBoldMathCommand\thetavec{\theta}
\DeclareBoldMathCommand\muvec{\mu}
\DeclareBoldMathCommand\tauvec{\tau}
\DeclareBoldMathCommand\nuvec{\nu}

\DeclareBoldMathCommand\Dmat{\mathrm{D}}
\DeclareBoldMathCommand\Ymat{\mathrm{Y}}
\DeclareBoldMathCommand\Imat{\mathrm{I}}
\DeclareBoldMathCommand\Kmat{\mathrm{Q}}
\DeclareBoldMathCommand\Kmat{\mathrm{K}}
\DeclareBoldMathCommand\Xmat{\mathrm{X}}
\DeclareBoldMathCommand\Zeromat{0}

\newcommand{\jax}{\pkg{JAX}}
\newcommand{\liesel}{\pkg{Liesel}}
\newcommand{\lieselgam}{\pkg{Liesel-GAM}}

\newcommand{\goose}{\pkg{Liesel-Goose}}
\newcommand{\lieselmodel}{\pkg{Liesel-Model}}
\newcommand{\pymc}{\pkg{PyMC}}

\newcommand{\stan}{\pkg{Stan}}

\newcommand{\tfplong}{\pkg{TensorFlow Probability}}
\newcommand{\tfpshort}{\pkg{TFP}}

\begin{document}

\maketitle

\clearpage

\section{Introduction}\label{sec:intro}

The implementation of Bayesian models is shaped by a recurring tension between several desirable goals. Researchers need software that supports the construction of complex hierarchical models while enabling both flexible and efficient posterior computation through modern tools such as automatic differentiation, just-in-time compilation, and hardware acceleration. They also need software that allows sophisticated model components to be reused across a range of applications.
These capabilities are often not equally well supported within a single software framework. This is particularly true for generalized additive regression models, which encompass an important class of reusable model components but are often better supported in specialized software than in general probabilistic programming frameworks.
As a result, methodological work often requires moving between software ecosystems and sacrificing either modeling convenience, inferential flexibility, or implementation efficiency.

In this space, \liesel{} is a \proglang{Python}-based probabilistic programming framework designed to reduce friction. It combines graph-based model construction, modular Markov chain Monte Carlo (MCMC) estimation, and dedicated support for generalized additive models (GAMs). The name refers to the central ``Gänseliesel'' fountain \footnote{English: Goose-Lizzy; Liesel is a colloquial version of the name Elisabeth.} in Göttingen, Germany, where \liesel{} was initially developed.

The framework consists of three parts. The \lieselmodel{} module provides a general library for expressing statistical models as directed acyclic graphs based on  Bayesian networks \citep{Koller2010-ProbabilisticGraphicalModels}. This representation supports stepwise model construction, a natural treatment of conditional dependencies, interactive probing and modification of model components, graph visualization, and graph-based tasks such as prediction and prior or posterior predictive simulation. Probability distributions and variable transformations are provided through the \tfplong{} (\tfpshort) interface \shortcites{Dillon2017} \citep{Dillon2017}. The \goose{} module provides a modular MCMC framework based on reusable kernels, including support for blocked componentwise sampling and for user-defined Gibbs and Metropolis-Hastings (MH) updates, while relying on \jax{} \shortcites{Bradbury2022} \citep{Bradbury2022} for automatic differentiation, just-in-time compilation, and accelerator support. It includes several popular general-purpose samplers ready for use, for example, Hamiltonian Monte Carlo (HMC) and the No-U-Turn Sampler (NUTS) via \pkg{BlackJAX}  \citep{Lao2022} and Iteratively Weighted Least Squares (IWLS) Metropolis-Hastings proposals. \lieselgam{} is a separate \proglang{Python} package that builds on \liesel{} and adds high-level building blocks for generalized additive models in \liesel{} by internally interfacing with \proglang{R} to obtain basis and penalty matrices from \pkg{mgcv} \citep{Wood2017}. Here, ``generalized additive models'' is used in a broad sense: the provided building blocks can be used not only for the mean as in traditional GAMs, but for any model parameter, enabling the specification of  generalized additive models for location, scale, and shape \citep[GAMLSS, ][]{Rigby2005} and developments beyond this model class. Out of the box, \lieselgam{} supports all effect types commonly used in generalized additive models, including linear effects, penalized splines, tensor product interactions, and spatial effects. It also provides convenient functionality for summaries and effect visualization.

This combination of ideas places \liesel{} in the context of  important strands of existing statistical software.
Among general probabilistic programming frameworks, \pkg{Stan} \shortcites{Carpenter2017-StanProbabilisticProgramming} \citep{Carpenter2017-StanProbabilisticProgramming, StanReferenceManual2-38} has set a standard for efficient inference in differentiable Bayesian models and has had a profound impact on applied Bayesian statistics through its robust implementation of Hamiltonian Monte Carlo and the No-U-Turn Sampler. Its design, however, is centered on this class of algorithms; it does not directly sample discrete parameters and is not primarily intended for user-defined componentwise MCMC schemes. \pkg{PyMC} \citep{Salvatier2016} provides a powerful and flexible probabilistic programming environment in \proglang{Python}, with broad accessibility and support for multiple step methods, including compound samplers. Compared with \liesel{}, its workflow is less centered on explicit graph-level model manipulation, and it does not provide reusable building blocks for generalized additive models. The \proglang{R} package \pkg{NIMBLE} \shortcites{deValpine2017} \citep{deValpine2017} is perhaps the closest relative in spirit. It is a highly capable and adaptable system with a \pkg{BUGS}-based declarative model specification language that is compiled into a graph representation and offers programmable algorithms, compilation for speed, and user-defined MCMC samplers. \liesel{} differs in focusing more directly on user-facing graph-based model specification, linking it closely to reusable semi-parametric regression components, and orienting its functionality in the \proglang{Python}/\jax{} ecosystem.

A different and equally important group of related software is focused primarily on generalized additive models, mostly implemented in \proglang{R}. In this space, \pkg{mgcv} \citep{Wood2017} is a foundational package for generalized additive modeling and remains a benchmark for the breadth, maturity, and computational quality of smooth-term methodology and implementation. The packages \pkg{gamlss} \citep{Stasinopoulos2017} and \pkg{bamlss} \citep{Umlauf2021} extend this perspective to more general distributional regression settings, with \pkg{bamlss} in particular providing a rich Bayesian and modular framework for generalized additive models. \citet{Buerkner2017} covers a wide class of Bayesian regression models with \pkg{brms}, accessible through a high-level formula interface, while delegating posterior computation to \pkg{Stan}. Finally, \pkg{BayesX} \citep{Brezger2005} is a highly developed system for Bayesian generalized additive models with substantial methodological breadth and efficient implementations.
The packages discussed above have fundamentally shaped the practice of Bayesian computation and generalized additive modeling, and the role of \liesel{} is not to duplicate their achievements or act as a mere replacement, but as a complementary framework with a different emphasis: facilitating methodological developments by providing graph-based modeling, modular MCMC development, and reusable generalized additive model components in a \proglang{Python}-native workflow.

The remainder of this paper is structured as follows. In Section~\ref{sec:example}, we briefly illustrate the full \liesel{} workflow by modeling the probability of detecting different species of collembola in German forests. In Sections~\ref{sec:liesel-model},~\ref{sec:goose}, and~\ref{sec:liesel-gam}, we provide details about the three parts of the framework, \lieselmodel{}, \goose{}, and \lieselgam{}, respectively, and use the example model from Section~\ref{sec:example} to illustrate additional details. These three sections are all split into a ``fundamental concepts'' and an ``advanced concepts'' subsection. In Section~\ref{sec:case-study-1}, we expand the application from Section~\ref{sec:example} into a more sophisticated ecological model to assess a biodiversity measure. In Section~\ref{sec:case-study-2}, we exploit \goose{} to contrast different MCMC sampling strategies in a generalized additive distributional regression model of childhood malnutrition in Zambia.

\liesel{} is available on the Python Package Index \footnote{\url{https://pypi.org/project/liesel/}} and is hosted and developed on GitHub \footnote{\url{https://github.com/liesel-devs/liesel}}. Documentation is available at \url{https://liesel-project.org}. Replication code, data, and output files for this paper are available at \url{https://github.com/liesel-devs/liesel-jss}.

\section{Motivating example} \label{sec:example}

In this section, we model the detection of different collembola species (also known as springtails) in forests in Lower Saxony, a state in northwest Germany. The example is intentionally chosen as a relatively standard Bayesian logistic regression model to provide a light overview of the \liesel{} workflow and introduce some of the central functionality. We will work with and modify the model created here throughout the paper and expand on this example in case study I to develop a more realistic occupancy model that can be used to assess common measures of species richness.

\subsection{Collembola in forests in Lower Saxony, Germany}

\liesel{} was previously used to analyze this data by \citet{Riebl2023}. The dataset is a subset of data collected at eight field sites (each comprising five plots) by the Research Training Group 2300, which is based at the University of Göttingen. The group investigates the ecosystem functions of pure and mixed forest stands comprising European beech, Norway spruce, and Douglas fir \shortcites{Glatthorn2023} \citep{Glatthorn2023}. The data is used to study the abundance and diversity of various taxa relevant to ecosystem functioning, including 26 species of collembola, which are tiny soil-dwelling hexapods that play an important role in decomposing organic matter and cycling nutrients. For this example, we consider a simplified analysis, focusing on the binary variable $\mathtt{detection}_{ij}$, which indicates whether any individuals of species $j = 1, \dots, M$ were found on experimental plot $i = 1, \dots, N$, where $M=26$ and $N=40$.

\subsection{Model setup} \label{sec:example-model}

We start by loading the data with

\begin{CodeChunk}
\begin{CodeInput}
>>> import pandas as pd
>>> collembola = pd.read_csv("data/collembola.csv")
\end{CodeInput}
\end{CodeChunk}

and importing the relevant core \liesel{} modules with

\begin{CodeChunk}
\begin{CodeInput}
>>> import liesel.model as lsl
>>> import liesel.goose as gs
\end{CodeInput}
\end{CodeChunk}

We also import the add-on library \lieselgam{} for functionality specific to generalized additive models
\begin{CodeChunk}
\begin{CodeInput}
>>> import liesel_gam as gam
\end{CodeInput}
\end{CodeChunk}

The data is organized in long format with $NM = 1040$ total observations. We set up a binomial logit model for $\mathtt{detection}_{ij} \sim \operatorname{Bernoulli}(\psi_{ij})$ with $\operatorname{logit}(\psi_{ij}) = \eta_i + \gamma_j$, where $\eta_i$ is an additive predictor and $\gamma_j$ is a species-specific random intercept.
We first initialize $\eta_i$ and add linear effects of the area potentially available (APA) for Norway spruce and Douglas fir:

\begin{CodeChunk}
\begin{CodeInput}
>>> tb = gam.TermBuilder.from_df(collembola)
>>> eta = gam.AdditivePredictor(name=r"$\eta$", intercept=True)
>>> eta += tb.lin("apa_spruce + apa_douglas")
\end{CodeInput}
\end{CodeChunk}

The \code{gam.TermBuilder} can be used to initialize a wide range of additive effect terms. In the \code{lin} constructor, linear terms can be defined using the right-hand side of Wilkinson formulas with a great deal of similarity to common \proglang{R} formula syntax.
We next add a spatial effect with a fixed-range Gaussian process covariance structure, also known as kriging, via

\begin{CodeChunk}
\begin{CodeInput}
>>> eta += tb.kriging(
...     "lon",
...     "lat",
...     k=40,
...     kernel_name="matern1.5",
...     linear_trend=False,
...     range=0.008047973,
...     inference=gs.MCMCSpec(gs.NUTSKernel),
... )
\end{CodeInput}
\end{CodeChunk}

Supplying \code{linear_trend=False} ensures stationarity.
The range parameter is chosen to let spatial correlation effectively decay to zero at a distance of about five kilometers, as used by \citet{Riebl2023}.
The \code{inference} argument attaches an MCMC specification to this term object, directing \goose{} to sample the corresponding parameters with a No-U-Turn Sampler
 \citep{Hoffman2014}; kernel specifications are described in detail in Section~\ref{sec:goose}.
The default used by \code{gam.TermBuilder} is a Metropolis-Hastings sampler with Iteratively Weighted Least Squares (IWLS) proposals \citep{Gamerman1997}. The argument \code{k} defines the number of knots to be used.
The species-specific random intercepts can be initialized with

\begin{CodeChunk}
\begin{CodeInput}
>>> gamma = tb.ri(
...     "species", name=r"$\gamma$", inference=gs.MCMCSpec(gs.NUTSKernel)
... )
\end{CodeInput}
\end{CodeChunk}
which, on implementation level, creates an $(M \times 1)$-dimensional coefficient object called $\beta_\gamma \sim \mathcal{N}(\mathbf{0}_M, \tau^2_\gamma \mathbf{I}_M)$ and a $(NM \times 1)$-dimensional object $\gamma$ that indexes $\beta_\gamma$ to align its elements with the data, which is stored in long format with $NM$ rows.

We set up $\operatorname{logit}(\psi_{ij})$ via

\begin{CodeChunk}
\begin{CodeInput}
>>> logit_psi = lsl.Var.new_calc(
...     lambda predictor, species_intercept: predictor + species_intercept,
...     predictor=eta,
...     species_intercept=gamma,
...     name=r"$logit(\psi)$",
... )
\end{CodeInput}
\end{CodeChunk}

The \code{lsl.Var} class is the basic building block of all \liesel{} models, and the \code{lsl.Var.new\_calc} constructor can be used to initialize a new variable as a deterministic, user-defined function of other variables, in this case a simple sum. The term constructors of \code{gam.TermBuilder} also return (somewhat specialized) instances of \code{lsl.Var}.

In the next step, we initialize the response variable using the \code{lsl.Var.new\_obs} constructor. The response's distribution is defined by passing a \code{lsl.Dist} that wraps a \tfplong{} (\tfpshort{}) distribution class. In this case, \code{tfd.Bernoulli} accepts its parameter directly on logit level, so no further transformation is required.

\begin{CodeChunk}
\begin{CodeInput}
>>> import tensorflow_probability.substrates.jax.distributions as tfd
>>> detection = lsl.Var.new_obs(
...     value=collembola["presence"],
...     dist=lsl.Dist(tfd.Bernoulli, logits=logit_psi),
...     name="detection",
... )
\end{CodeInput}
\end{CodeChunk}

\begin{figure}[tb]
    \centering
    \includegraphics[width=\linewidth, trim={5em 4.3em 5em 4em}, clip]{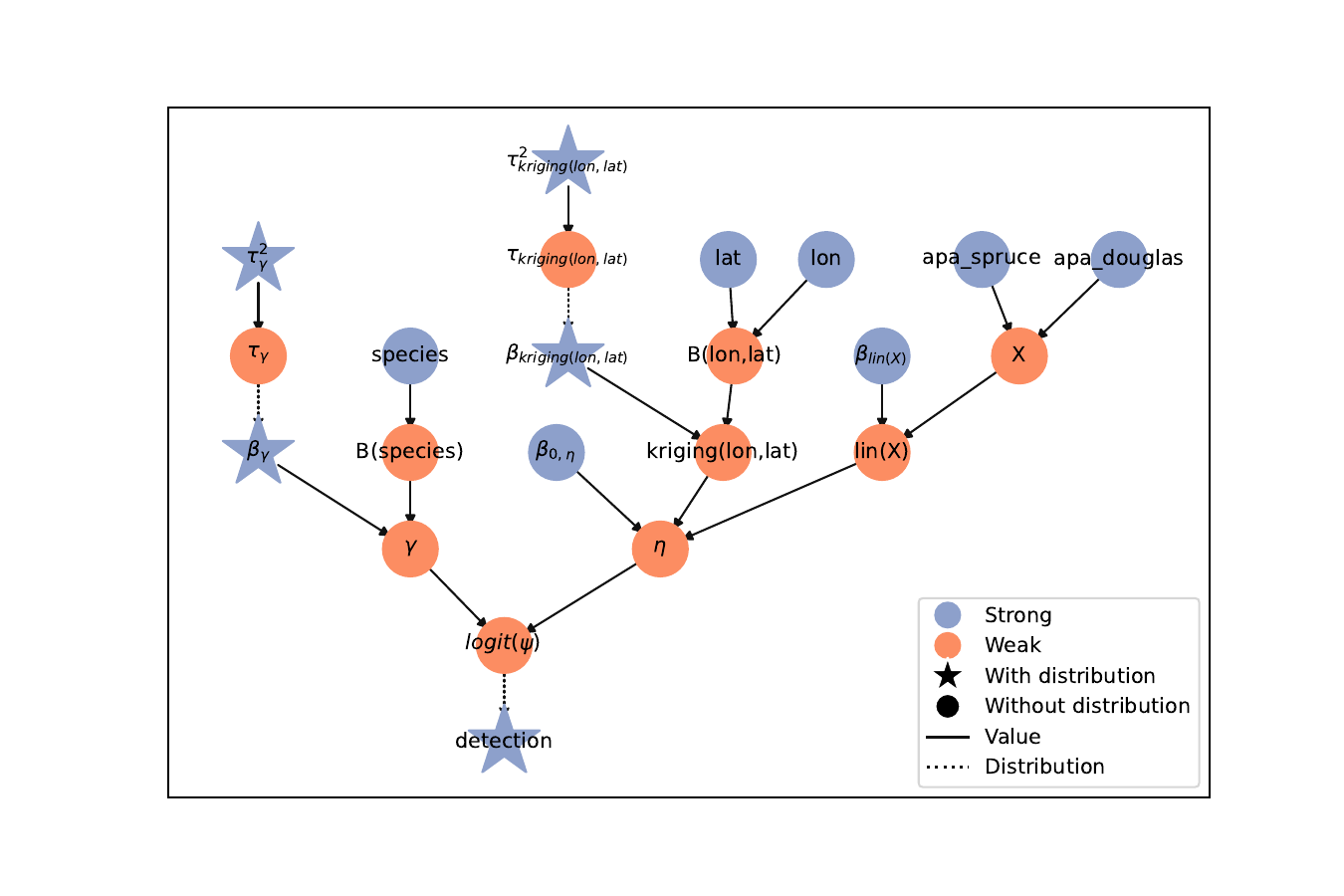}
    \caption{Model graph for the spatial logistic regression model of collembola detection  via \code{lsl.Model.plot()}. Similar plots of the subgraphs for individual variables can be created via \code{lsl.Var.plot()}.}
    \label{fig:a1-model}
\end{figure}

We now initialize the model and plot the model graph, which is included in Figure~\ref{fig:a1-model}.
\begin{CodeChunk}
\begin{CodeInput}
>>> model = lsl.Model(detection)
>>> model.plot()
\end{CodeInput}
\end{CodeChunk}
The graph shows connections between the model variables. Note that both the random intercept and the spatial effect are represented in a basis function framework with bases $B(\text{species})$ and $B(\text{lon}, \text{lat})$, and corresponding coefficients $\beta$, respectively. These coefficients are set up with multivariate normal prior distributions by \code{gam.TermBuilder} with covariance structures according to their effect type, see Section~\ref{sec:liesel-gam} for details. The graph in Figure~\ref{fig:a1-model} further reveals that additional random variables  were created by the \code{gam.TermBuilder} constructors during model setup, namely the variance parameters $\tau^2_\gamma$ and $\tau^2_{\text{kriging(lon,lat)}}$. By default, they are initialized with inverse Gamma priors. While users are usually expected to retrieve this information from the package documentation, it can also be revealed by interactively inspecting the variable through
\begin{Code}
>>> model.vars[r"$\tau_{\gamma}^2$"].dist_node.distribution.__name__
'InverseGamma'
\end{Code}
Exploiting conditional conjugacy, the default variance parameters are initialized with Gibbs sampler setup information in their \code{lsl.Var.inference} attributes (see Section~\ref{sec:liesel-gam}).

\subsection{MCMC sampling and diagnostics} \label{sec:example-mcmc}

For MCMC sampling, \goose{} uses the specifications stored in \code{lsl.Var.inference} to construct a component-wise sampling algorithm. It can be run by calling

\begin{figure}[tb]
    \centering
    \begin{minipage}{.4\linewidth}
        \includegraphics[width=\linewidth, trim={0 0 7em 0}, clip]{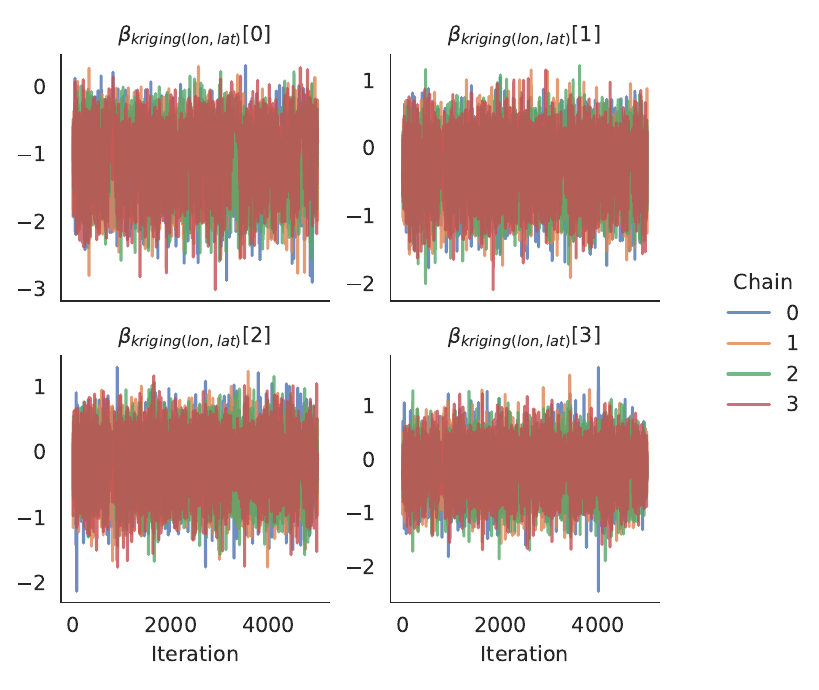}
    \end{minipage}
    \begin{minipage}{.59\linewidth}
        \includegraphics[width=\linewidth]{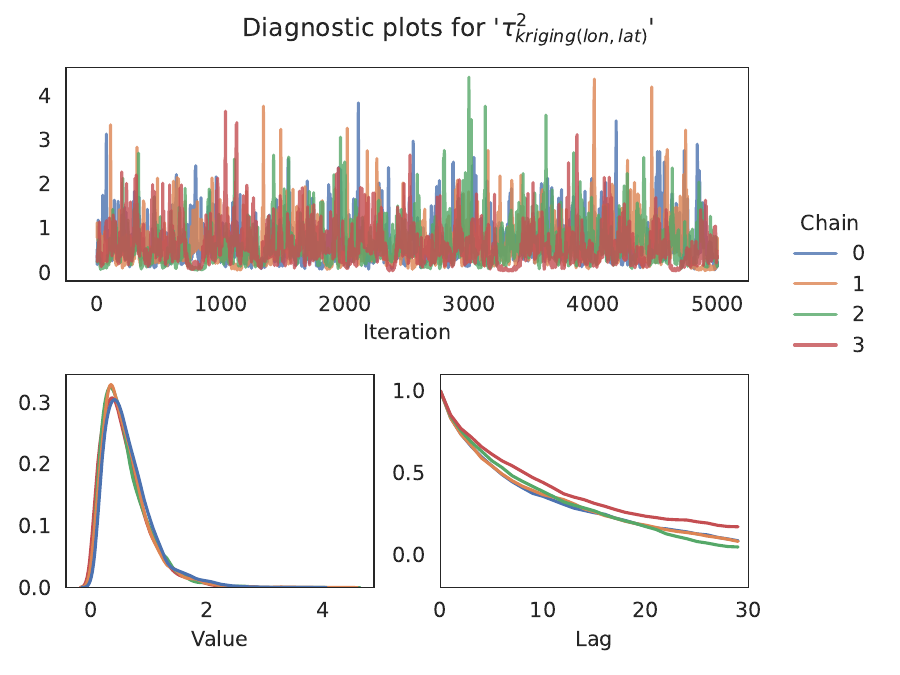}
    \end{minipage}
    \caption{Left: Trace plots for the first four elements of $\beta_{kriging(lon,lat)}$ via \code{gs.plot\_trace()}. Right: Extended parameter diagnostics of $\tau_{kriging(lon,lat)}^2$ via
    \code{gs.plot\_param()}. By default, these plots include only posterior samples.
    }
    \label{fig:a1_trace_and_diag}
\end{figure}

\begin{CodeChunk}
\begin{CodeInput}
>>> results = gs.LieselMCMC(model).run_for_epochs(
...     seed=314, num_chains=4, adaptation=2000, posterior=5000
... )
\end{CodeInput}
\begin{CodeOutput}
liesel.goose.engine - INFO - Initializing kernels...
liesel.goose.engine - INFO - Done
liesel.goose.engine - INFO - Starting epoch: FAST_ADAPTATION, 200 transitions
liesel.goose.engine - WARNING - Errors per chain for kernel_00: 2, 1, 2, 3
liesel.goose.engine - WARNING - Errors per chain for kernel_01: 3, 3, 4, 2
liesel.goose.engine - INFO - Finished epoch
... [truncated]
liesel.goose.engine - INFO - Starting epoch: POSTERIOR, 5000 transitions
liesel.goose.engine - WARNING - Errors per chain for kernel_00: 0, 1, 0, 0
liesel.goose.engine - INFO - Finished epoch
\end{CodeOutput}
\end{CodeChunk}
The \goose{} sampling engine prints a log while sampling, keeping the user informed about progress and the occurrence of errors during sampling. After sampling, the errors are included in the summary constructed via \code{gs.Summary(results)}. In this case, we observe a small number of divergent transitions in the NUTS kernels for $\beta_{kriging(lon,lat)}$ and $\boldsymbol{\beta}_\gamma$, mostly during warmup.
An excerpt of the posterior summary (without error messages) of the parametric coefficients and the variance parameters is included in Table~\ref{tab:a1-summary}.
\begin{table}[tb]
    \centering
    \small
\begin{tabular}{llrrrrrrrr}
\toprule
Variable & Index & Mean & SD & ESS (bulk) & ESS (tail) & $\hat R$ & q .05 & q .5 & q .95 \\
\midrule
$\beta_{0,\eta}$ & () & -1.86 & 0.41 & 336.00 & 572.29 & 1.01 & -2.54 & -1.86 & -1.18 \\
$\beta_{lin(X)}$ & (0,) & 0.02 & 0.36 & 1238.59 & 2349.35 & 1.00 & -0.57 & 0.03 & 0.59 \\
$\beta_{lin(X)}$ & (1,) & -0.22 & 0.37 & 1123.21 & 2420.36 & 1.00 & -0.83 & -0.21 & 0.37 \\
$\tau_{\gamma}^2$ & () & 3.36 & 1.14 & 7742.00 & 12495.07 & 1.00 & 1.91 & 3.16 & 5.45 \\
$\tau_{kriging(lon,lat)}^2$ & () & 0.61 & 0.42 & 428.85 & 358.59 & 1.02 & 0.13 & 0.51 & 1.41 \\
\bottomrule
\end{tabular}
    \caption{Excerpt from the posterior parameter summary table for the collembola detection model as generated by \goose{} via \code{gs.Summary(results).to\_dataframe()}. Rounded to two decimal places. In the linear coefficient $\beta_{lin(X)}$, the first element corresponds to \texttt{apa\_spruce} and the second to \texttt{apa\_douglas}.
    \label{tab:a1-summary}
    }
\end{table}
A brief overview of central diagnostics is available through the summary with
\begin{CodeChunk}
\begin{CodeInput}
>>> summary = gs.Summary(results)
>>> print(summary.aggregate_diagnostics().iloc[:, :3])
\end{CodeInput}
\begin{CodeOutput}
                                ess_bulk      ess_tail      rhat
parameter                                                       
$\beta_{0,\eta}$              336.000753    572.289326  1.008055
$\beta_{\gamma}$              719.301971   1394.129998  1.003043
$\beta_{kriging(lon,lat)}$   1031.786729   1023.204127  1.006576
$\beta_{lin(X)}$             1123.212316   2349.348432  1.003260
$\tau_{\gamma}^2$            7742.000867  12495.071918  1.001102
$\tau_{kriging(lon,lat)}^2$   428.853176    358.591426  1.016659
\end{CodeOutput}
\end{CodeChunk}
For the multivariate parameters, such as $\beta_{\gamma}$, this aggregation includes the minimum effective sample size and the maximum $\hat{R}$ to give a worst-case summary. Visual diagnostic checks can be performed using several \code{gs.plot\_x} functions, where \code{x} can be \code{trace} for trace plots, \code{cor} for autocorrelation plots, \code{pairs} for pairwise scatter plots, \code{density} for posterior kernel density estimates (kde), or \code{param} for a triplet of trace plot, autocorrelation plot, and kde. Figure~\ref{fig:a1_trace_and_diag} shows the output of
\begin{CodeChunk}
\begin{CodeInput}
>>> gs.plot_trace(
...     results,
...     params=[r"$\beta_{kriging(lon,lat)}$"],
...     param_indices=[0, 1, 2, 3]
... )
>>> gs.plot_param(results, param=r"$\tau_{kriging(lon,lat)}^2$")
\end{CodeInput}
\end{CodeChunk}

\subsection{Summaries, visualizations, and prediction} \label{sec:example-results}
\begin{figure}[tb]
    \centering
    \begin{minipage}{.49\linewidth}
        \includegraphics[width=\linewidth]{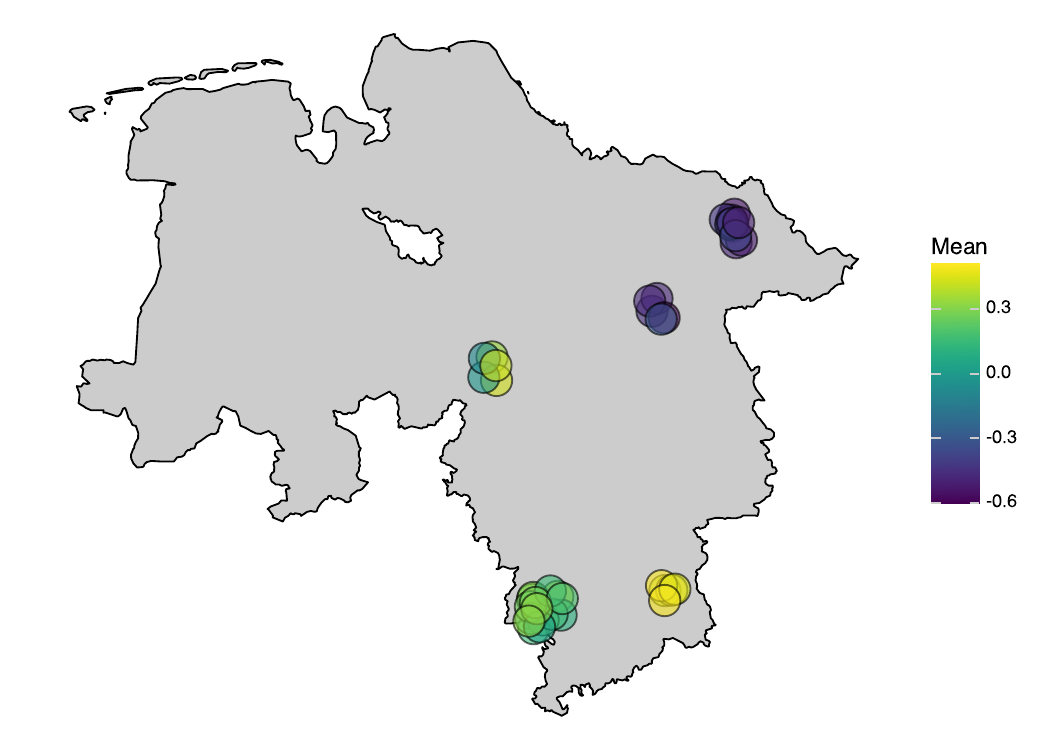}
    \end{minipage}
    \begin{minipage}{.49\linewidth}
        \includegraphics[width=\linewidth]{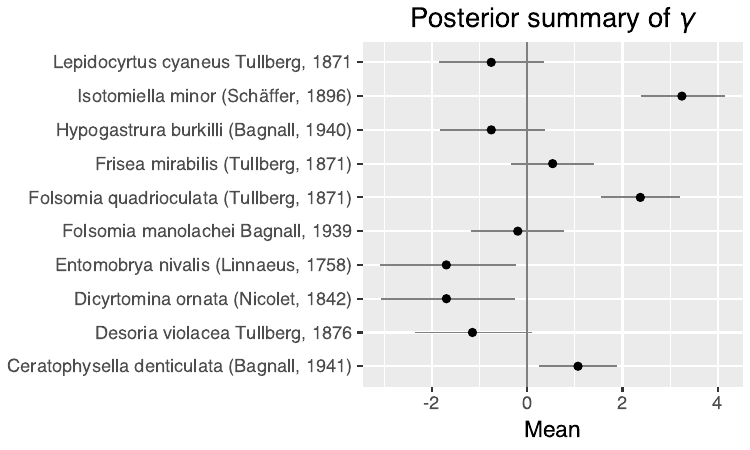}
    \end{minipage}
    \caption{Left: Posterior mean of the spatial effect, evaluated at the 40 experimental plots, obtained by manually plotting the result of \code{lsl.Model.predict()}. Right: Posterior mean and 90\% credible intervals of the random intercept for the first ten species of collembola, obtained via \code{gam.plot\_forest()}.}
    \label{fig:a1-results}
\end{figure}

A dictionary of posterior samples for all sampled parameters can be extracted with
\begin{CodeChunk}
\begin{CodeInput}
>>> samples = results.get_posterior_samples()
\end{CodeInput}
\end{CodeChunk}
and used for posterior prediction in \code{lsl.Model.predict()}:
\begin{CodeChunk}
\begin{CodeInput}
>>> predictions = model.predict(samples, newdata=None)
\end{CodeInput}
\end{CodeChunk}
The argument \code{newdata} can be used to pass a dictionary of values for model variables at which to evaluate the predictions; the default \code{newdata=None}, which is used here, simply takes the current values of the model variables. The predictions are returned as arrays of shape \code{(nchains, nsamples, n)}, where \code{n} is the dimension of the variable. In this case, the predictions for the spatial effect have shape
\begin{Code}
>>> predictions["kriging(lon,lat)"].shape
(4, 5000, 1040)
\end{Code}
since we used four chains, 5000 samples per chain, and the variable has shape $NM=1040$. The left panel in Figure~\ref{fig:a1-results} shows the posterior mean of the spatial effect at the experimental plots included in the study. The plot suggests that, on average, the probability of finding collembola in forest soil is higher in the plots located in southern and central Lower Saxony.

\lieselgam{} provides additional functionality for quickly plotting posterior summaries of common additive effect types. The right panel in Figure~\ref{fig:a1-results} shows a forest plot of the intercept for the first ten species of collembola with 90\% credible intervals, generated with
\begin{CodeChunk}
\begin{CodeInput}
>>> gam.plot_forest(gamma, samples, ci_quantiles=(0.05, 0.95))
\end{CodeInput}
\end{CodeChunk}
The plot suggests notable inter-species differences, with, for example, Isotomiella minor occurring more frequently and Entomobrya nivalis occurring less frequently than average.

\section[Liesel-Model]{\lieselmodel{}}
\label{sec:liesel-model}

\subsection{Introduction} \label{sec:model-intro}

\lieselmodel{} is a module in \liesel{} that allows users to construct and modify Bayesian models as directed acyclic graphs. The graph is used to efficiently evaluate complex hierarchical joint log density or probability functions, to compute predictions, to draw prior or posterior predictive samples, and for visualizations of the model structure.
We import the module as
\begin{Code}
>>> import liesel.model as lsl
\end{Code}

\subsection{Fundamental concepts} \label{sec:model-fundamentals}

\subsubsection[The Liesel model graph]{The \liesel{} model graph}

\liesel{} models are directed acyclic graphs in which nodes represent variables $\mathcal{X} = \{X_1, \dots, X_n\}$ with values (or realizations) $\bm{x} = (x_1, \dots, x_n)$, and edges encode statistical and deterministic dependencies. They can be regarded as an extended version of Bayesian networks \citep{Koller2010-ProbabilisticGraphicalModels}.
The goal is to define the joint log density $\log p(\bm{x})$, which decomposes over the graph structure.
\liesel{} distinguishes two kinds of variables according to how their value $x_i$ is determined.
Strong variables, forming the set $\mathcal{S} \subseteq \mathcal{X}$, have their value supplied externally.
Weak variables, forming the set $\mathcal{W} = \mathcal{X} \setminus \mathcal{S}$, have their value  computed as a function of other variables in the graph, i.e., they satisfy $X_i = f_i(\mathcal{P}^V_i),$ where $f_i$ is a deterministic function and $\mathcal{P}^V_i \subseteq \mathcal{X}$ denotes what we call the value-parents of $X_i$.

Each variable may additionally be assigned a distribution, whose parameters are supplied by a set of distribution-parents $\mathcal{P}^D_i \subseteq \mathcal{X}$.
The joint log density of all model variables $\mathcal{X}$ with realizations $\boldsymbol{x}$ then decomposes as
\begin{equation}
    \label{eq:logprob}
    \log p(\boldsymbol{x}) = \sum_{i=1}^n \log p(x_i \mid \mathcal{P}^D_i),
\end{equation}
where variables without a distribution contribute zero by convention.
Figure~\ref{fig:a1-model} shows an example model graph with strong and weak nodes distinguished by color and where dependencies are indicated by directed edges with different line styles for type of dependence.

The set of strong variables is further partitioned into observed variables $\mathcal{Y} \subseteq \mathcal{S}$, parameter variables $\mathcal{Z} \subseteq \mathcal{S}$, and constant variables $\mathcal{C} = \mathcal{S} \setminus (\mathcal{Y} \cup \mathcal{Z})$. Table~\ref{tab:model-symbols} gives an overview of the various variable subsets.
Observed variables may or may not carry a distribution: response variables carry a distribution that defines the likelihood, while covariates and other data inputs often do not.
The distribution of parameter variables defines their prior.
The weak variables in $\mathcal{W}$ represent intermediate deterministic quantities such as transformed parameters or regression predictors, and typically do not carry a distribution.

The unnormalized log posterior can be written in terms of the realizations of the observed variables, $\boldsymbol{y} = (x_i)_{i \in \mathcal{I}_{\mathcal{Y}}}$, and the realizations of the parameter variables, $\boldsymbol{z} = (x_i)_{i \in \mathcal{I}_{\mathcal{Z}}}$, as
\begin{equation} \label{eq:logprob-decomposition}
    \log p(\boldsymbol{z} \mid \boldsymbol{y})
	\propto \log p(\boldsymbol{x})
	= \sum_{i \in \mathcal{I}_{\mathcal{Y}}} \log p(x_i \mid \mathcal{P}^D_{i})
		+  \sum_{i \in \mathcal{I}_{\mathcal{Z}}} \log p(x_i \mid \mathcal{P}^D_{i}),
\end{equation}
where $\mathcal{I}_\mathcal{Y}$ is the set of indices of the observed variables in $\mathcal{Y}$; index sets for the other variable classes are analogously defined (see also Table~\ref{tab:model-symbols}).
In the very common case that no observed variable is among the parents of any parameter variable, \eqref{eq:logprob-decomposition} can be understood as the classical decomposition into likelihood and prior, i.e.,
\begin{equation}
    \log p(\boldsymbol{z} \mid \boldsymbol{y}) \propto \log p(\boldsymbol{y} \mid \boldsymbol{z})  + \log p(\boldsymbol{z}).
\end{equation}

In some models it is useful to assign a proper distribution to a weak variable. Then, the log density $\log p(x_i \mid \mathcal{P}^D_i)$ is evaluated at the propagated value $x_i = f_i(\mathcal{P}^V_i)$ (see Section~\ref{sec:model-advanced}).
This arises naturally in copula models where a joint distribution is decomposed into univariate marginals and a copula in accordance with Sklar's theorem \citep{Sklar1959-FonctionsRepartitionDimensions}.

\begin{table}[t]
\small
    \centering
    \begin{tabularx}{\linewidth}{lllX}
\toprule
   Symbol  & Elements & Values & Description \\
   \midrule
   $\mathcal{I}$ & $\{1, \dots, n\}$ & - &  Set of the indices of all variables in a model. \\
   $\mathcal{X}$ & $\{X_i : i \in \mathcal{I}\}$ &
   $\boldsymbol{x} := (x_i)_{i \in \mathcal{I}}$ & All variables. \\
   $\mathcal{S} \subseteq \mathcal{X} $ & $\{X_i : i \in \mathcal{I}_{\mathcal{S}}\}$ & $\boldsymbol{s} := (x_i)_{i \in \mathcal{I}_\mathcal{S}}$ & Strong variables. \\
   $\mathcal{W} = \mathcal{X} \setminus \mathcal{S}$ & $\{X_i : i \in \mathcal{I}_{\mathcal{W}}\}$ & $\boldsymbol{w} := (x_i)_{i \in \mathcal{I}_\mathcal{W}}$ & Weak variables. \\
   \addlinespace
   $\mathcal{Y} \subseteq \mathcal{S}$ & $\{X_i : i \in \mathcal{I}_{\mathcal{Y}}\}$ & $\boldsymbol{y} := (x_i)_{i \in \mathcal{I}_\mathcal{Y}}$ & Observed variables. \\
   $\mathcal{Z} \subseteq \mathcal{S}\setminus \mathcal{Y}$ & $\{X_i : i \in \mathcal{I}_{\mathcal{Z}}\}$ & $\boldsymbol{z} := (x_i)_{i \in \mathcal{I}_\mathcal{Z}}$ & Parameter (latent) variables. \\
   $\mathcal{C} = \mathcal{S} \setminus (\mathcal{Y} \cup \mathcal{Z})$ & $\{X_i : i \in \mathcal{I}_{\mathcal{C}}\}$ & $\boldsymbol{c} := (x_i)_{i \in \mathcal{I}_\mathcal{C}}$ & Constant variables. \\
 \bottomrule
\end{tabularx}
\caption{Overview of variable subsets and realizations in a \liesel{} model. For any subset $\mathcal{A}$ of variables, we denote the subset of indices corresponding to that subset by $\mathcal{I}_{\mathcal{A}}$. \label{tab:model-symbols}}
\end{table}

\subsubsection[The Var class]{The \code{Var} class}

The \code{lsl.Var} class represents random variables and provides four constructors that cover the main use cases. They are described below.

\begin{description}
    \item[\code{lsl.Var.new\_param()}] Creates a parameter
          variable with an associated prior distribution.
          In many cases, the values of parameter variables are the targets of estimation in
          a \liesel{} model. The argument \code{inference} is used to store information on how inference for this variable should be performed. For example \code{gs.MCMCSpec} provides information on MCMC sampling, see Section~\ref{sec:goose}.

    \item[\code{lsl.Var.new\_obs()}] Creates an observed variable. The value is fixed to the observed data, and the distribution specifies the likelihood.

    \item[\code{lsl.Var.new\_calc()}] Creates a weak variable whose value is deterministically computed from other variables in the model via \code{function}. Usually, no distribution is associated with a weak variable, although exceptions are possible (see advanced topics). The constructor accepts self-written \proglang{Python} functions. To exploit \jax{}'s just-in-time compilation and automatic differentiation, these functions must be pure, i.e., have no side-effects (see below for further details). Weak variables generally keep a cache and only re-compute their value if at least one of their inputs are outdated. Caching can be turned off by passing \code{cache=False}, since in some cases, it may be more beneficial to save memory by allowing repeated evaluations of cheap functions.

    \item[\code{lsl.Var.new\_value()}] Creates a constant variable.
\end{description}

The (optional) argument \code{name} can be used to pass a unique identifier,
which is used for graph visualizations and for lookup operations on the \code{lsl.Model}
class. If no name is passed, a unique name will be automatically generated when a model
is initialized.
For all constructors that accept the \code{dist} argument, it is possible to pass
\code{dist=None}, which creates a
nonrandom variable. In the case of \code{lsl.Var.new\_param}, this is equivalent to a parameter with a
constant prior distribution. An overview table of the most important attributes and methods of the \code{lsl.Var} class is included as Table~\ref{tab:var-and-dist} in Appendix~\ref{app:model-details}.

Variables always know their inputs and expose them through \code{lsl.Var.all\_input\_vars()}. After being added to a \code{lsl.Model}, variables also know their outputs, which are available via \code{lsl.Var.all\_output\_vars()}. Table~\ref{tab:var-access} additionally details how individual distribution- and value-inputs to a variable can be accessed and replaced via square-bracket syntax resembling the corresponding operations on \proglang{Python} dictionaries and lists.

\begin{table}[tb]
    \centering
    \footnotesize
    \begin{tabularx}{\linewidth}{lX}
         \toprule
         Code & Description \\
         \midrule
         \multicolumn{2}{l}{\textbf{\textsf{Value inputs}}} \\
         \addlinespace
         \multicolumn{2}{l}{\code{v1 = lsl.Var.new\_calc(jnp.exp, x=y)}} \\
         \multicolumn{2}{l}{\code{v2 = lsl.Var.new\_calc(jnp.exp, y)}} \\
         \addlinespace
         \code{v1.value\_node["x"]} & Access keyword argument \code{"x"} to the function defining the value of \code{v1}.\\
         \code{v1.value\_node["x"] = z} & Replace the input to the keyword argument \code{"x"} with the variable \code{z}.\\
         \addlinespace
         \code{v2.value\_node[0]} & Access first positional argument to the function defining the value of \code{v2}.\\
         \code{v2.value\_node[0] = z} & Replace the input to the first positional argument with the variable \code{z}.\\
         \midrule
         \multicolumn{2}{l}{\textbf{\textsf{Distribution inputs}}} \\
         \addlinespace
         \multicolumn{2}{l}{\code{v3 = lsl.Var.new\_param(..., lsl.Dist(tfd.Normal, loc=y, scale=1.0))}} \\
         \multicolumn{2}{l}{\code{v4 = lsl.Var.new\_param(..., lsl.Dist(tfd.Normal, y, 1.0))}} \\
         \addlinespace
         \code{v3.dist\_node["loc"]} & Access keyword argument \code{"loc"} to the distribution of \code{v3}.\\
         \code{v3.dist\_node["loc"] = z} & Replace the input to the keyword argument \code{"loc"} with the variable \code{z}.\\
         \addlinespace
         \code{v4.dist\_node[0]} & Access first positional argument to the distribution of \code{v4}.\\
         \code{v4.dist\_node[0] = z} & Replace the input to the first positional argument with the variable \code{z}.\\
         \addlinespace
         \code{v4.dist\_node = lsl.Dist(...)} & Completely replace the distribution of \code{v4} with a different distribution.\\
         \code{v4.dist\_node = None} & Remove the distribution of \code{v4}, turning it into a nonrandom variable.\\
         \bottomrule
    \end{tabularx}
    \caption{Access to and replacement of inputs to \liesel{} variables and their distributions.\label{tab:var-access}}
\end{table}

\subsubsection[The Dist class]{The \code{Dist} class}

The \code{Dist} class represents a random variable's probability distribution by
wrapping a distribution class that follows the \tfpshort{} interface.
The simplest case is passing a distribution class with fixed parameters.
\begin{CodeChunk}
\begin{CodeInput}
>>> lsl.Dist(tfd.Normal, loc=0.0, scale=1.0)
\end{CodeInput}
\end{CodeChunk}
When distribution parameters depend on other variables, they are passed to the
distribution.
\begin{CodeChunk}
\begin{CodeInput}
>>> s = lsl.Var.new_param(1.0, name="sigma")
>>> lsl.Dist(tfd.Normal, loc=0.0, scale=s)
\end{CodeInput}
\end{CodeChunk}
Any parameters accepted by the wrapped \tfpshort{} \code{Distribution} must be passed to the \code{lsl.Dist} class instead; in this case, the parameters are called \code{loc} and \code{scale}.
The \code{lsl.Dist} class also accepts self-written distribution classes or simple
callables with return values that follow the \tfpshort{} distribution interface.

\subsubsection{Variable transformations}

It is often useful to map continuous random variables with a constrained domain to the full real
line using bijective functions. \liesel{} allows for such variable transformations through
the method \code{lsl.Var.biject()}, which applies a bijection to the variable itself,
and \code{Dist.biject_parameters()}, which applies bijections to the parameters of
the distribution. In both cases, the model is reparameterized using the multivariate
change of variables theorem: Let $X$ denote a $d$-dimensional continuous
random variable with domain
$\mathcal{H} \subseteq \mathbb{R}^d$, and let $h: \mathcal{H} \rightarrow \mathbb{R}^d$
denote a bijective, continuously differentiable link function with continuously differentiable inverse $h^{-1}$. Then the density of the transformed variable
$\widetilde X = h(X)$ is given by
\begin{equation}
    \label{eq:bijection}
    \tilde{p}(\tilde{x}) = p(h^{-1}(\tilde{x})) \left| \operatorname{det}(J) \right|,
\end{equation}
where $J$ is the Jacobian of $h^{-1}$. In \liesel{}, variable transformations are applied
by passing a bijector object following the \tfpshort{}  \code{Bijector} interface, which
offers a wide range of common bijectors ready for use.
For most builtin continuous distributions, \tfpshort{} defines default bijectors that
map from the distribution's domain to the reals. These defaults can be used by
passing \code{bijector="auto"} to \code{lsl.Var.biject()}, which is convenient but sacrifices some transparency about the specific bijector being used. During transformation of a
variable $X$, a new variable $\widetilde{X}=h(X)$ with density given by \eqref{eq:bijection} is
created, and the variable $X$ is turned into a weak variable without distribution and with value
$x = h^{-1}(\tilde{x})$. The status of $X$ as a parameter or observed node is transferred to $\widetilde{X}$ and set to \code{False} on $X$.

Using the collembola detection model from Section~\ref{sec:example}, a log-transformation of the spatial effect's variance parameter can be implemented with
\begin{CodeChunk}
\begin{CodeInput}
>>> import tensorflow_probability.substrates.jax.bijectors as tfb
>>> tau2_spatial = model.copy().vars[r"$\tau_{kriging(lon,lat)}^2$"]
...
>>> tau2_spatial.biject(tfb.Exp(), inference=gs.MCMCSpec(gs.NUTSKernel))
\end{CodeInput}
\end{CodeChunk}
We copy the model here to avoid modifying the original model.
We also supply a new MCMC specification to the \code{inference} argument to account for the fact that the existing specification on \code{tau2\_spatial} is outdated after transformation. The specification is applied to the transformed variable and removed from the original one:
\begin{CodeChunk}
\begin{Code}
>>> print(tau2_spatial.inference)
None
>>> print(tau2_spatial.bijected_var.inference)
MCMCSpec(<class 'liesel.goose.nuts.NUTSKernel'>, self.kernel_group=None)
\end{Code}
\end{CodeChunk}

A variable with existing inference specification cannot be transformed without explicit handling of the inference attribute: the existing specification must be removed either by simply passing \code{inference="drop"} or by passing an inference object for the transformed parameter, as we do here.

\subsubsection[The Model class]{The \code{Model} class}
The \code{lsl.Model} class represents a Bayesian network of \code{lsl.Var} objects as a directed acyclic
graph. It can be initialized by passing a list of nodes. It is sufficient to pass leaf nodes, i.e., nodes without outputs. From these leaf nodes, the model class infers the full graph upon
initialization by recursing through each node's inputs and stores a topological ordering of the nodes. Note that, different from common \proglang{R} model classes, the \code{lsl.Model} does not perform a model fit upon initialization. The main purposes of the \code{lsl.Model} class are:
\begin{enumerate}
    \tightlist
    \item Compute the model's joint probability $p(\boldsymbol{x})$ and update it when the
value of a node in the graph changes. The joint probability is available through \code{lsl.Model.log_prob}.
The model also provides the attributes \code{lsl.Model.log_lik} and \code{lsl.Model.log_prior}.
    \item Plot the model graph through \code{lsl.Model.plot()}.
    \item Compute predictions at new observed values through \code{lsl.Model.predict()}.
    \item Hierarchically draw prior or posterior predictive samples for all or a subset of the random nodes in the model through \code{lsl.Model.sample()}. Note that this method does not draw from the posterior distribution with density $p(\boldsymbol{z} \mid \boldsymbol{y})$, but from the joint distribution with density $p(\boldsymbol{x})$. Sampling from the posterior is described in Section~\ref{sec:goose}.
    \item Access variables and subsets of variables with \code{lsl.Model.vars}, \code{lsl.Model.parameters} and \code{lsl.Model.observed}.
    \item Modify an existing model with \code{lsl.Model.add()}, \code{lsl.Model.replace()} and\\ \code{lsl.Model.join()}.
\end{enumerate}
An overview table of the most important attributes and methods of the \code{lsl.Model} class is included as Table~\ref{tab:model} in Appendix~\ref{app:model-details}.

\subsubsection{Updating the model state}

To avoid inefficient recomputation of all computed quantities in the model graph
whenever any input changes, each node in the model keeps track of its direct child nodes (its outputs) and
maintains a flag that indicates whether its own cached value or log probability are outdated. When a node receives a new value, it sets the \code{outdated} flag to \code{True} for all nodes in its recursive output set.
An update of all outdated nodes can be performed by calling \code{lsl.Model.update()}, which is detailed in Algorithm~\ref{alg:model-update}. The method iterates over all nodes in the model and checks their \code{outdated} flags.
If the flag is \code{False}, the cached value is returned immediately.
If the flag is \code{True}, the node recomputes its value, updates the cache, and sets the flag to \code{False}.
This process recurses through the input nodes as needed and ensures that only outdated
computations are re-evaluated;
and such re-evaluations occur only once for each change of the model state. This efficiency gain is particularly important for large models with many parameters.

Given a \code{position} dictionary of variable names and values, the model can perform a pure update (i.e., without side-effects) via \code{lsl.Model.update\_state()}. Instead of modifying the model state inplace, this method returns an updated model state object and leaves the current model state untouched.
Below, using the collembola detection model from Section~\ref{sec:example}, we first print the model's unnormalized log probability at the initial values, then update the model state with the posterior means, override the model state with the updated state, and again print the log probability to verify that the state has been updated:
\begin{CodeChunk}
\begin{Code}
>>> model.log_prob
Array(-755.37256, dtype=float32)

>>> model.state = model.update_state(position=summary.quantities["mean"])
>>> model.log_prob
Array(-379.41083, dtype=float32)
\end{Code}
\end{CodeChunk}

\begin{algorithm}[t]
\caption{\code{lsl.Model.update()}.\label{alg:model-update}}
\footnotesize
\begin{algorithmic}[1]
\State Let $X_{(1)}, \ldots, X_{(n)}$ be a topological ordering of the model variables.
\For{$i = 1, \ldots, n$}
    \If{$X_{(i)}$ is outdated}
        \State Re-evaluate the value and/or conditional log probability of $X_{(i)}$.
    \EndIf
\EndFor
\end{algorithmic}
\end{algorithm}

\subsubsection{Predictions and samples}

The model graph can be used to compute predictions via \code{lsl.Model.predict()} given new values $\boldsymbol{y}^*$ and $\boldsymbol{c}^*$ for
(subsets of) the observed and constant variables in $\mathcal{Y}$ and $\mathcal{C}$, respectively, passed as \code{newdata}. The method additionally accepts a
collection of values
$\boldsymbol{z}^{[1]}, \dots, \boldsymbol{z}^{[T]}$
the parameter
variables in $\mathcal{Z}$; which is usually a collection of posterior samples and passed as \code{samples}. First, the model state is updated once by setting the values of the observed nodes to the new data supplied in $\boldsymbol{y}^*$. Next, for each $t = 1, \dots, T$, the values of the parameter variables are set to the values supplied in $\boldsymbol{z}^{[t]}$, the model state is updated using Algorithm~\ref{alg:model-update}, and the values of all model variables are recorded in $\boldsymbol{x}^{*[t]}$. The function returns the collection of predicted model states $\boldsymbol{x}^{*[1]}, \dots, \boldsymbol{x}^{*[T]}$ for all nodes, but the set of nodes for which predictions are computed and returned can be narrowed by passing a list of names to \code{lsl.Model.predict()}:

\begin{CodeChunk}
\begin{Code}
>>> newdata = {
...     "apa_spruce": jnp.array([0.5, 0.6]),
...     "apa_douglas": jnp.array([0.0, 0.0]),
... }
>>> lin_predicted = model.predict(
...     samples,
...     predict=["lin(X)"],
...     newdata=newdata,
... )

>>> list(lin_predicted)
['lin(X)']

>>> lin_predicted["lin(X)"].shape
(4, 5000, 2)
\end{Code}
\end{CodeChunk}

In a similar way, samples from the model (not to be confused with samples from the parameters' posterior distribution) can be drawn via \code{lsl.Model.sample()}, which traverses the model graph in topological order and draws pseudo-random samples from the distributions of the model's random variables given the current model state:

\begin{CodeChunk}
\begin{CodeInput}
>>> from jax.random import key
>>> prior_pred_samples = model.sample(shape=(), seed=key(1535))
>>> list(prior_pred_samples)
\end{CodeInput}
\begin{CodeOutput}
['$\\beta_{\\gamma}$',
 '$\\beta_{kriging(lon,lat)}$',
 '$\\tau_{\\gamma}^2$',
 '$\\tau_{kriging(lon,lat)}^2$',
 'z']
\end{CodeOutput}
\begin{Code}
>>> prior_pred_samples["detection"].shape
(1040,)
\end{Code}
\end{CodeChunk}

If a collection of posterior samples $\boldsymbol{z}^{[1]}, \dots, \boldsymbol{z}^{[T]}$ for the parameter variables $\mathcal{Z}$ (or a subset thereof) is supplied, \code{lsl.Model.sample()} produces samples from the posterior predictive distribution of the observed random variables in the model:

\begin{CodeChunk}
\begin{Code}
>>> posterior_pred_samples = model.sample(
...     shape=(), seed=key(1538), posterior_samples=samples
... )
>>> posterior_pred_samples["detection"].shape
(4, 5000, 1040)
\end{Code}
\end{CodeChunk}

If no values are supplied for $\boldsymbol{z}$, it produces hierarchical samples from the prior predictive distribution of all random variables in the model. Note that, in this case, no samples are drawn for parameter variables with constant priors; they are held constant at their current values. Pseudocode for \code{lsl.Model.sample()} is given in Algorithm~\ref{alg:model-sample}.

Note that, to keep the exposition concise, we use an overly strict notation for the allowed values passed to \code{newdata} and \code{samples}. In practice, values for any strong variable in the model, including parameters and constants, can be supplied via \code{newdata} and \code{samples}, as long as no variable is present in both.

\begin{algorithm}[tb]
\caption{\code{lsl.Model.sample()}.\label{alg:model-sample}}
\footnotesize
\begin{algorithmic}[1]
\State Let $X_{(1)}, \ldots, X_{(n)}$ be a topological ordering of the model variables.
\If{\code{newdata} is not None}
    \State Update the model state with $\boldsymbol{y}^*$ and $\boldsymbol{c}^*$ via Algorithm~\ref{alg:model-update}.
\EndIf
\For{$t = 1, \ldots, T$}
    \State $j \gets 0$
    \For{$i = 1, \ldots, n$}
        \If{a value for $X_{(i)}$ is given in $\boldsymbol{z}^{[t]}$}
            \State Set the value of $X_{(i)}$ to the corresponding value in $\boldsymbol{z}^{[t]}$.
        \ElsIf{$X_{(i)}$ is has a non-degenerate distribution}
            \State Update the model state via Algorithm~\ref{alg:model-update}.
            \State Sample a new value of $X_{(i)}$ from its conditional distribution given the current state of its parents.
            \State $j \gets j + 1$
            \State Record the value of $X_{(i)}$ as $\tilde{\boldsymbol{x}}^{[t]}_j$.
        \Else
            \State \textbf{skip}
        \EndIf
    \EndFor
    \State $J \gets j$
    \State Set $\tilde{\boldsymbol{x}}^{[t]} \gets (\tilde{\boldsymbol{x}}^{[t]}_1, \ldots, \tilde{\boldsymbol{x}}^{[t]}_J)$.
\EndFor
\State \Return Collection of sampled values $\tilde{\boldsymbol{x}}^{[1]}, \ldots, \tilde{\boldsymbol{x}}^{[T]}$
\end{algorithmic}
\end{algorithm}

\subsection{Advanced topics} \label{sec:model-advanced}

\subsubsection{Statistical vs. computational graph}

Above, we introduced a simplified version of \liesel{}'s model graph that treats a random variable's value and log probability as two attributes of the same node (the random variable). We call this the \textit{statistical} graph. This representation is concise, it is useful in most use cases, and it corresponds to common depictions of statistical models as graphs.

The implementation, however, treats a variable's value and log probability as two separate nodes, and performs operations on the basis of a \textit{computational} graph. In the computational graph, all nodes represent a single (albeit possibly multi-dimensional) value, and all edges represent deterministic input-output relationships. The basic building block of the computational graph is the abstract \code{lsl.Node} class with its subclasses \code{lsl.Value} (representing constants), \code{lsl.Calc} (representing computations), and \code{lsl.Dist}, which can be interpreted as a specialized \code{lsl.Calc}. A \code{lsl.Var} always wraps a value-node (either \code{lsl.Value} or \code{lsl.Calc}), and optionally wraps an additional \code{lsl.Dist} node. The computational graph can be plotted via \code{lsl.Model.plot\_nodes()}, and the value and dist node are available as attributes on the \code{lsl.Var}. We expect most users of \liesel{} to work exclusively with the statistical graph.

\subsubsection[Pure functions for JAX compatibility]{Pure functions for \jax{} compatibility}

\jax{}'s just-in-time compilation and automatic differentiation require pure functions.
A pure function's output depends only on its inputs and has no side effects.
However, the \lieselmodel{} graph is not pure in this sense, since each node maintains internal state (the cached value and outdated flag).
\liesel{} resolves this by treating the graph's state explicitly as input and output of functions derived from the graph, predominantly building on the function \code{lsl.Model.update\_state()} described above.

Additionally, all functions used in \code{lsl.Var} objects must be \jax{}-compatible. This means, for example, that users cannot use global mutable state inside functions supplied to calculator variables \code{lsl.Var.new\_calc()}, and that
\code{jax.numpy} and \code{jax.scipy} must be used instead of \code{numpy} and \code{scipy} functionality. This transition is facilitated by the fact that the \jax{} interfaces are almost exact mirrors of the \code{numpy} and \code{scipy} interfaces. Existing code can therefore often be ported with limited effort, and knowledge from working with the ubiquituous \code{numpy} and \code{scipy} libraries carries over (with some exceptions noted in the \jax{} documentation).
Additionally, users cannot use builtin
\proglang{Python} control flow (e.g., \code{if} statements), for-loops or while-loops and must instead use the alternatives provided by \jax{}. We provide a brief overview in Table~\ref{tab:jax-api}.

Functions that are not repeatedly evaluated during model fitting, such as basis function evaluations for fixed covariate values, are exempt from these restrictions. \liesel{} recognizes that the corresponding nodes never become outdated and simply uses their cached values.

\begin{table}[t]
    \centering
    \footnotesize
    \begin{tabularx}{\linewidth}{lXX}
    \toprule
         Non-\jax{}& \jax{} alternative & Description  \\
         \midrule
         \code{numpy} & \code{jax.numpy} & Vectorized computations with arrays.\\
         \code{scipy} & \code{jax.scipy} & General functions for scientific computing.\\
         \code{If: ...} & \code{jax.lax.cond()}, \allowbreak\ \code{jax.lax.select()} &Control flow.\\
         \code{For: ... } & \code{jax.lax.fori\_loop()}, \allowbreak\ \code{jax.lax.scan()} & For-loops.\\
         \code{While: ... } & \code{jax.lax.while\_loop()} &While-loops.\\
         \bottomrule
    \end{tabularx}
    \caption{Overview of pure, JIT- and autodiff-compatible \jax{} alternatives to common \proglang{Python} functions.\label{tab:jax-api}}
\end{table}

\subsubsection{Debugging}

Debugging a dysfunctional model is facilitated by the interactive nature of \liesel{} models. Specifically, \code{lsl.Model} and \code{lsl.Var} objects can be probed in a \proglang{Python} REPL (read-evaluate-print-loop) and in Jupyter notebooks. The model's structure can be assessed by inspecting the model graph via \code{lsl.Model.plot()} and \code{lsl.Var.plot()}, and problematic variables, for example variables with \code{NaN} or \code{Inf} values or log probabilities, can be identified by inspecting diagnostic dataframes returned by \code{lsl.Model.diagnose()} and \code{lsl.Var.diagnose()}.

\section[Liesel-Goose]{\goose{}}
\label{sec:goose}

\subsection{Introduction} \label{sec:goose-intro}

\goose{} is the MCMC module in \liesel{}. It provides general infrastructure for component-wise Markov chain Monte Carlo (MCMC) by combining MCMC kernels via composition \citep[]{Johnson2013-ComponentWiseMarkovChain,Brooks2011-HandbookMarkovChain}.
The library implements a modular general sampling loop, a selection of popular general-purpose MCMC kernels, step size tuning via the dual averaging algorithm \citep{Neal2006-OptimalScalingPartially,Hoffman2014}, and building blocks for the implementation of bespoke Gibbs and Metropolis-Hastings MCMC kernels.
We import the module as
\begin{Code}
>>> import liesel.goose as gs
\end{Code}

\subsection{Fundamental concepts} \label{sec:goose-fundamentals}

\subsubsection[Component-wise MCMC sampling in Liesel-Goose]{Component-wise MCMC sampling in \goose{}}

 The general sampling loop is outlined in pseudocode in Algorithm~\ref{alg:mcmc}, which is run in a user-defined number of parallel chains. In the sampling loop, the set of parameters $\mathcal{Z}$ is partitioned into $B$ ordered blocks $(\mathcal{Z}_b)$, $b = 1, \dots, B$, with values $\boldsymbol{z}_b = (x_i), i \in \mathcal{I}_{\mathcal{Z}_b}$. Each block is updated in a fixed sequence using an MCMC kernel $k_b$ that draws samples from the full conditional distribution of the selected parameters. The kernels are governed by hyperparameters denoted by $\psi_b$, which can be fixed manually or tuned during an adaptation phase. The algorithm is initialized at the current model state, but can include random perturbations (jittering), which is described below in Section~\ref{sec:goose-advanced}.

 \begin{algorithm}[bt]
\caption{Component-wise Markov chain Monte Carlo in \goose{}.\label{alg:mcmc}}
\footnotesize
\begin{algorithmic}[1]
\State Partition $\mathcal{Z}$ into $B$ ordered blocks $\mathcal{Z}_1, \ldots, \mathcal{Z}_B$.
\State For each block $b= 1, \dots, B$, let $\boldsymbol{z}_b = (x_i), i \in \mathcal{I}_{\mathcal{Z}_b}$ denote its values and let $k_b$ denote its MCMC kernel, with state $\psi_b^{[t]}$ at transition $t$.
\State Initialize $\boldsymbol{x}_1^{[1]}$ using the current model state or a jittered version thereof.
\For{$t = 1, \ldots, T$}
    \For{$b = 1, \ldots, B$}
        \State $\boldsymbol{z}_b^{[t]} \gets k_b(\boldsymbol{x}_b^{[t]}, \psi_b^{[t]})$ via \code{Kernel.transition()}.
        \State Update the model state to $\boldsymbol{x}_{b+1}^{[t]}$ via Algorithm~\ref{alg:model-update}.
    \EndFor
    \State $\boldsymbol{z}^{[t]} \gets (\boldsymbol{z}_1^{[t]}, \ldots, \boldsymbol{z}_B^{[t]})$.
    \State $\boldsymbol{x}_1^{[t+1]} \gets \boldsymbol{x}_{B+1}^{[t]}$.
\EndFor
\State \Return Collection of posterior samples $\boldsymbol{z}^{[1]}, \ldots, \boldsymbol{z}^{[T]}$
\end{algorithmic}
\end{algorithm}

\subsubsection{Sampling phases}
\begin{figure}[t]
    \centering
    \includegraphics[width=\linewidth, trim={0 5.5cm 0 1cm}, clip]{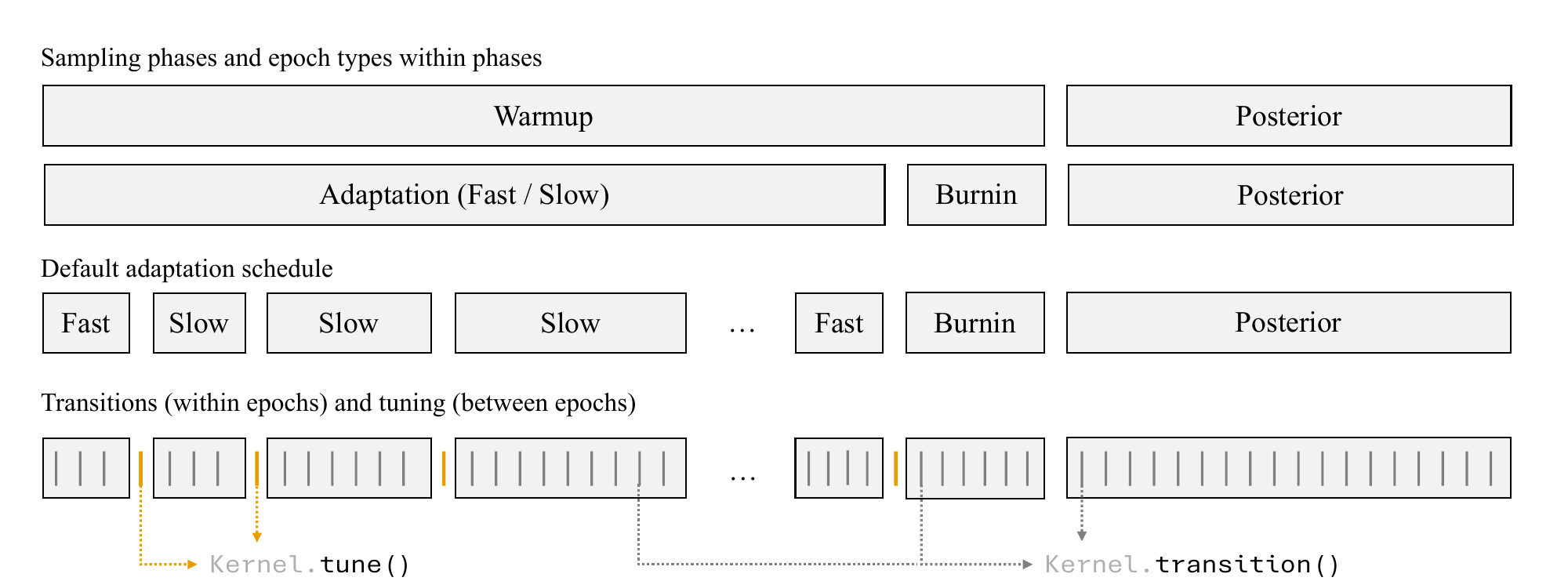}
    \caption{Overview of the phases in the \goose{} MCMC algorithm on different levels. The relative lengths of the warmup and posterior sampling phases are not representative of typical lengths. Usually, all warmup samples are discarded for inference. In the adaptation phase, the hyperparameters of MCMC kernels can be tuned within each transition, or in between epochs.}
    \label{fig:sampling-phases}
\end{figure}

Sampling proceeds for a fixed number of $T$ transitions, which are divided into a \textit{warmup} and a \textit{posterior} phase, and grouped into epochs within phases. An overview is provided in Figure~\ref{fig:sampling-phases}. The warmup phase distinguishes fast and slow adaptation and burnin epochs. Adaptation epochs must precede burnin, which must precede posterior epochs. Within the adaptation phase, kernel hyperparameter tuning can occur inside transitions or between epochs. The distinction of fast and slow adaptation epochs can be used to perform different tuning operations, enabling for example the implementation of a \pkg{Stan}-like hyperparameter tuning for Hamiltonian Monte Carlo kernels \citep{StanReferenceManual2-38} with several expanding slow-adaptation epochs sandwiched in between two fast-adaptation epochs. This schedule serves as the model for the default adaptation schedule in \goose{}.

Although conditions under which adaptive MCMC satisfies ergodicity have been explored \citep[see, for example,][]{Roberts2007-CouplingErgodicityAdaptive}, the main purpose of the adaptation phase in \goose{} is for the algorithm to reach the typical set of the posterior distribution while allowing MCMC kernels to update their hyperparameters. Samples from the adaptation phase are discarded together with burnin samples; inference is then based only on samples from the posterior phase.

\subsubsection{MCMC kernels}

\goose{} provides several MCMC kernels that can be used directly:

\begin{description}
\item[MHKernel.] The \code{MHKernel} implements a Metropolis-Hastings sampler as a wrapper around a user-defined function generating proposals based on the current model state. If the proposal distribution is asymmetric, the function must also return the Metropolis-Hastings correction factor. An optional step size argument is also provided, which can be tuned using the dual averaging algorithm to reach a user-defined target acceptance probability by scaling proposals.

\item[GibbsKernel.] The \code{GibbsKernel} wraps a user-defined function generating samples from a full conditional distribution into a \goose{}-compatible MCMC kernel. With a Gibbs kernel, no tuning is necessary or possible, since all proposals are accepted.

\item[RWKernel.] The \code{RWKernel} implements a Gaussian random walk proposal distribution with a Metropolis-Hastings acceptance step. The kernel uses the dual averaging algorithm to adjust the step size (i.e.~to scale the proposal distribution) during adaptation epochs, such that a user-defined target acceptance rate, which defaults to 0.234 \citep[see][]{Gelman1997}, is reached.

\item[HMCKernel and NUTSKernel.] The \code{HMCKernel} and \code{NUTSKernel} use the gradient of the log-posterior to generate MCMC chains with a low autocorrelation. The implementation wraps \pkg{BlackJAX}'s implementations of the Hamiltonian Monte Carlo \citep[HMC,][]{Neal2011} and No-U-Turn Sampler \shortcites{Lao2020} \citep[NUTS,][]{Hoffman2014, Lao2020, Phan2019} algorithms. Both kernels tune the step size during adaptation epochs using the dual averaging algorithm. After slow adaptation epochs, the mass matrix (or vector, if the default diagonal metric is used) of the momentum is set to the empirical covariance (variance) of the samples from the preceding slow epoch.

\item[IWLSKernel.] The \code{IWLSKernel} is named after the method proposed by \citet{Gamerman1997}, which is often used for Bayesian distributional regression models \citep{Brezger2005}. However, \liesel{}'s implementation is also inspired by the closely related Riemannian manifold Metropolis-adjusted Langevin algorithm  \citep[MMALA,][]{Girolami2011}. This approach allows us to add a step size parameter in a straightforward way, which can then be tuned with the dual averaging algorithm during fast and slow adaptation epochs. More precisely, the \code{IWLSKernel} employs a Gaussian proposal density and a Metropolis-Hastings correction, where the mean vector $\muvec$ and the covariance matrix $\boldsymbol{\Sigma}$ of the proposal distribution depend on the gradient (score) and the negative Hessian (Hess) of the log full conditional posterior via
$$
\muvec = \boldsymbol{z}_b - \nicefrac{s^2}{2} \operatorname{Hess}(\boldsymbol{z}_b)^{-1} \operatorname{score}(\boldsymbol{z}_b), \qquad \boldsymbol{\Sigma} = -s^2 \operatorname{Hess}(\boldsymbol{z}_b)^{-1},
$$
where $s$ denotes the step size and $\boldsymbol{z}_b$ the parameter vector. The factor $\nicefrac{1}{2}$ that is multiplied with $s^2$ in the mean vector arises from the Langevin diffusion, which is the basis of the MMALA algorithm.
The gradient and Hessian are computed via \jax{}'s automatic differentiation.
The default behavior of using the observed negative Hessian can be changed by the user by providing a function for computing the Cholesky decomposition of the negative Hessian from the model state. This can allow for model-specific improvements to efficiency and numerical stability, for example, when the expected negative Hessian is used instead of the observed \citep{Klein2015-BayesianStructuredAdditiveb}.
\end{description}

The default \goose{} kernels are sufficient to estimate many statistical models with MCMC. However, \goose{} was specifically designed for cases that require specialized kernels. In these situations, new kernel classes adhering to the \code{gs.Kernel} interface can be implemented. The developer does not need to start from scratch, however, since \goose{} is designed to offer building blocks that facilitate the implementation of new kernel classes. For example, if a kernel should support step size tuning via dual averaging, \goose{} can extend the kernel state with the necessary fields and offers functions for adjusting the step size. A mixin for Metropolis-Hastings kernels is provided as well. We provide further details below in Section~\ref{sec:goose-advanced}.

\subsubsection{Specification of kernels, blocks, and their order}
The default ordering of parameter blocks is based on a reversed topological sort, so that all parent nodes come after their children in the ordering. This choice is based on the observation that, in many models, observed variables holding the data to be modeled, such as the response in a regression model, appear late in the topological ordering. Defaulting to a reverse topological ordering corresponds to starting with those parameters that are closest to the data.
The default partition assigns each parameter variable with \code{lsl.Var.inference} not \code{None} to their own block, so $B = \lvert \mathcal{Z} \rvert$ if all parameters are equipped with an inference specification. Using \code{gs.MCMCSpec}, both the ordering and the block assignment can be defined by the user, for example
\begin{CodeChunk}
\begin{CodeInput}
>>> spec_a = gs.MCMCSpec(gs.NUTSKernel, kernel_group="a", order=2)
>>> spec_b = gs.MCMCSpec(gs.NUTSKernel, kernel_group="b", order=1)
>>> v1 = lsl.Var.new_param(1.0, inference=spec_a)
>>> v2 = lsl.Var.new_param(1.0, inference=spec_a)
>>> v3 = lsl.Var.new_param(1.0, inference=spec_b)
\end{CodeInput}
\end{CodeChunk}
will assign \code{v1} and \code{v2} to the same block and arrange the algorithm such that \code{v3} is sampled before \code{v1} and \code{v2}. Note that it is not sufficient to pass the same \code{gs.MCMCSpec} object to two variables to trigger grouping; grouping has to be done explicitly through the \code{kernel\_group} argument.

\subsubsection{Finding good initial values}

While initial values are theoretically irrelevant in the asymptotic limit, they can have a strong effect on the speed and quality of convergence in practice. \goose{} therefore includes a simple routine for gradient-based optimization that interfaces with the \jax{}-based optimization library \pkg{Optax} \shortcites{deepmind2020jax} \citep{deepmind2020jax}, which offers a wide range of popular optimizers such as Adam, RMSprop, and the limited-memory  Broyden-Fletcher-Goldfarb-Shanno algorithm (L-BFGS), a quasi-Newton optimizer. Optimization is available in \goose{} via \code{gs.optim\_flat()}. Below, we run the default optimization to find approximations to the posterior modes of the regression coefficients using Adam and update the model state with the results.

\begin{CodeChunk}
\begin{Code}
>>> params = [pname for pname in model.parameters if "tau" not in pname]
>>> optresult = gs.optim_flat(model, params=params)
Training loss: 388.426, Validation loss: 388.426:  1544/10000

>>> model.state = model.update_state(optresult.position)
\end{Code}
\end{CodeChunk}

We exclude the variance parameters from optimization here for simplicity, because enforcing their bounded domain would require optimization on a transformed level.
A validation model can be used for early stopping, if supplied. Otherwise, early stopping occurs when updates to the training loss become negligible.

\subsubsection{Summary and diagnostics}

Via \code{gs.Summary}, \goose{} offers a number of summary statistics for aggregating posterior samples and diagnosing the sampling process. The general summary includes the mean and standard deviation, posterior quantiles, and highest posterior density (HPD) intervals.

The included diagnostics are the rank-normalized effective sample size (ESS), the Gelman-Rubin statistic $\hat{R}$, and Markov chain standard errors (MCSE) \citep{Vehtari2021-RankNormalizationFoldingLocalizationa}. Additionally, acceptance probabilities can be assessed via \code{gs.Summary.acceptance\_probability\_df()},
and a summary of error codes for each MCMC kernel is provided in the default summary representation and in \code{gs.Summary.error\_df()}. The type and severity of error codes varies. Common examples include divergent transitions, which can occur in HMC and NUTS kernels and may indicate problematic posterior geometry and maximum tree depth errors, which may arise in NUTS kernels when no U-turn was observed within the maximum number of steps and may indicate an issue of efficiency rather than correctness \citep{Betancourt2018-ConceptualIntroductionHamiltonian}. Other common warnings include invalid transition probabilities (\code{NaN} or \code{Inf}), which can indicate numerical issues such as numerical overflow.

For complex models with high-dimensional parameters, we also offer the aggregation method \code{gs.Summary.aggregate\_diagnostics()}, which can be used to obtain the minimum ESS and maximum $\hat{R}$ for each sampled parameter, providing a compact worst-case summary.
Further details regarding the diagnostics can be found in \citet{Vehtari2021-RankNormalizationFoldingLocalizationa}. Highest posterior density intervals and diagnostics are computed via \pkg{ArviZ} \citep{kumar2019arviz}.

\subsubsection{Model choice}
Several common Bayesian model choice criteria represent efforts to estimate the expected log predictive density for new observations, i.e., the expected out-of-sample log likelihood. Such efforts include the Watanabe-Akaike/Widely-Applicable Information criterion \citep[WAIC,][]{Watanabe2010-AsymptoticEquivalenceBayes} and pareto-smoothed importance sampling estimates for leave-one-out cross-validation \citep[PSIS-LOO,][]{Vehtari2017-PracticalBayesianModel}. While WAIC and PSIS-LOO are asymptotically equivalent, \citet{Vehtari2017-PracticalBayesianModel} find PSIS-LOO to be more robust in finite data scenarios and generally recommend it over WAIC. Both WAIC and PSIS-LOO require pointwise log likelihood evaluations, which can be obtained from a \liesel{} model and a samples dictionary via \code{lsl.log\_prob\_pointwise(model.observed, samples)}. \goose{} then provides \code{gs.loo(lpp, samples)}, which uses the pointwise log probability evaluations to compute PSIS-LOO via \pkg{ArviZ} \citep{kumar2019arviz}.

\subsection{Advanced topics} \label{sec:goose-advanced}

\subsubsection{Dual averaging}

For step size tuning, \goose{} uses the dual averaging algorithm
\citep{Nesterov2009} in the notation of \citet{Hoffman2014}. All builtin MCMC kernels that use dual averaging accept the algorithm's hyperparameters, which are summarized in Table~\ref{tab:dual-averaging}, but we want to point out the guidance by the \pkg{Stan} development team to not change the defaults without experience with the algorithm \citep[, Chapter 15.2]{StanReferenceManual2-38}. The most promising hyperparameter to set manually is the target acceptance probability, which may be increased or decreased to nudge the algorithm towards larger or smaller step sizes, respectively.

\begin{table}[t]
\small
    \centering
    \begin{tabularx}{\linewidth}{lllX}
    \toprule
         Parameter & Domain & Default & Description \\
         \midrule
         \code{initial\_step\_size} & $(0, \infty)$ & Varies & Initial step size. \\
         \code{da\_target\_accept} & $[0,1]$ & Varies & Target acceptance probability.\\
         \code{da\_gamma} & $(0, \infty)$ &$0.05$& The adaptation regularization scale. Higher values mean smaller updates.\\
         \code{da\_kappa} & $(0, \infty)$ &$0.75$& The adaptation relaxation exponent. Controls, how much earlier iterations affect current step sizes, higher values mean smaller influence of earlier iterations.\\
         \code{da\_t0} &$(0, \infty)$ &$10$& Dampens early exploration: a large value prevents early updates from being very large.\\
         \bottomrule
    \end{tabularx}
    \caption{Arguments for controlling the hyperparameters of the dual averaging algorithm used in several MCMC kernels in \goose{}.\label{tab:dual-averaging}}
\end{table}

For using dual averaging in custom MCMC kernels, \goose{} provides a module available as \code{liesel.goose.da} that includes a kernel state protocol (\code{DAKernelState}), a function for initializing a dual averaging kernel state (\code{da\_init}), a function that updates a kernel state by performing a single dual averaging algorithm step (\code{da\_step}), and a function for finalizing a kernel state after conclusion of the algorithm (\code{da\_finalize}).

\subsubsection{Setting up a custom Gibbs sampler}

In the collembola detection model from Section~\ref{sec:example}, the full conditional distribution for the random intercept's variance $\tau^2_\gamma$ is an inverse gamma distribution $\tau^2_\gamma \mid \cdot  \sim\operatorname{InvGamma}(\tilde a, \tilde b)$ with $\tilde a = a + \nicefrac{1}{2}M$ and $\tilde b = b + \nicefrac{1}{2}\boldsymbol{\beta}_{\gamma}^\top \boldsymbol{\beta}_\gamma$, where $a$ and $b$ are the concentration and scale parameters of the parameter's inverse gamma prior. In \goose{}, the transition function for this Gibbs sampler can be defined with
\begin{CodeChunk}
\begin{Code}
>>> coef_name = r"$\beta_{\gamma}$"
>>> M = model.vars[coef_name].value.size

>>> tau2_gamma = model.vars[r"$\tau_{\gamma}^2$"]
>>> a_prior = tau2_gamma.dist_node["concentration"].value
>>> b_prior = tau2_gamma.dist_node["scale"].value

>>> def tau2_transition(prng_key, model_state):
...     position = model.extract_position([coef_name], model_state)
...     coef_current = position[coef_name]
...     a_gibbs = a_prior + 0.5 * M
...     b_gibbs = b_prior + 0.5 * (coef_current.T @ coef_current)
...     dist = tfd.InverseGamma(concentration=a_gibbs, scale=b_gibbs)
...     tau2_gibbs = dist.sample(sample_shape=(), seed=prng_key)
...     return {tau2_gamma.name: tau2_gibbs}
\end{Code}
\end{CodeChunk}

Every Gibbs transition function must accept the arguments \code{prng\_key} (the seed for pseudo-random number generation, PRNG) and \code{model\_state}. The \code{prng\_key} is required to allow \goose{} to explicitly handle the state of the pseudo-random number generation, thus keeping the function pure and compatible with \jax{}'s just-in-time compilation.
To finalize the Gibbs sampling setup, we use the \code{gs.GibbsKernel.with_transition_fn} classmethod to define a small factory function that constructs a full \code{gs.GibbsKernel} instance, and pass it to the variable's \code{inference} attribute:

\begin{CodeChunk}
\begin{Code}
>>> tau2_gamma.inference = gs.MCMCSpec(
...     gs.GibbsKernel.with_transition_fn(tau2_transition)
... ) 
\end{Code}
\end{CodeChunk}

\subsubsection{Custom MCMC kernels}

The easiest way to use a custom MCMC kernel in \goose{} is to provide a proposal function for a \code{gs.MHKernel}. The function must accept a pseudo-random number key, a model state and a step size as arguments, and
be compatible with just-in-time compilation via \jax{} (i.e., pure, without side-effects). It returns a \code{gs.MHProposal}, which simply wraps the proposed value and the Metropolis-Hastings log-correction factor $\log p(\boldsymbol{z}_b^{[t-1]} | \boldsymbol{z}_b^*) - \log p(\boldsymbol{z}_b^* | \boldsymbol{z}_b^{[t-1]})$.
The \code{gs.MHKernel} handles the acceptance/rejection logic and is fully equipped with dual averaging functionality for step size tuning, which can be switched on by passing \code{da_tune_step_size} as a keyword argument to the kernel. In this case, users should ensure that their settings for the initial step size (default: $1$) and the target acceptance probability (default: $.234$) are suitable.

A random walk kernel for the linear regression coefficients in the collembola presence model from Section~\ref{sec:example} can be implemented with

\begin{CodeChunk}
\begin{Code}
>>> coef_name = model.vars["lin(X)"].coef.name
>>> def rw_proposal(prng_key, model_state, step_size):
...     pos = model.extract_position([coef_name], model_state)
...     current = pos[coef_name]
...
...     proposal_dist = tfd.Normal(loc=current, scale=step_size)
...     proposed = proposal_dist.sample(seed=prng_key)
...
...     backward_dist = tfd.Normal(loc=proposed, scale=step_size)
...     backward_log_prob = backward_dist.log_prob(current)
...     forward_log_prob = proposal_dist.log_prob(proposed)
...     log_correction = (backward_log_prob - forward_log_prob).sum()
...     return gs.MHProposal({coef_name: proposed}, log_correction)
\end{Code}
\end{CodeChunk}
It can then be attached to the coefficient variable with
\begin{CodeChunk}
\begin{Code}
>>> model.vars["lin(X)"].coef.inference = gs.MCMCSpec(
...     gs.MHKernel,
...     kernel_kwargs={"proposal_fn": rw_proposal, "da_tune_step_size": True},
... )
\end{Code}
\end{CodeChunk}
In this case, the proposal distribution is symmetric, so the log correction factor is zero by definition. We still compute it here explicitly for the purpose of demonstration.

While a custom proposal function for a \code{gs.MHKernel} can be written conveniently, it may not cover cases in which a custom MCMC kernel requires additional hyperparameters or specialized tuning. For such cases, \goose{} provides tools for users to write their own classes, implementing the \code{gs.Kernel} protocol. We provide a step-by-step tutorial in the online documentation.\footnote{\url{https://docs.liesel-project.org/en/latest/tutorials/md/08-custom-kernel.html\#fully-customized-mcmc-kernel}}

\subsubsection{Jittering start values}

Jittering start values in MCMC sampling refers to initializing multiple chains at dispersed points in the parameter space--typically by adding small random perturbations to a common baseline. This practice can help diagnose convergence issues by revealing whether chains with different initial values mix toward the same stationary distribution. Some MCMC frameworks use random initialization by default; \pkg{Stan}, for example, draws starting values from $\operatorname{Unif}(-2, 2)$ in the unconstrained parameter space. \liesel{} does not perform random initialization by default, but offers functionality for fine-grained control of random initialization through arguments
of \code{gs.MCMCSpec}:
\begin{CodeChunk}
\begin{Code}
>>> beta_lin = model.vars[r"$\beta_{lin(X)}$"]
>>> beta_lin.inference = gs.MCMCSpec(
...     gs.NUTSKernel,
...     jitter_dist=tfd.Uniform(low=-2.0, high=2.0),
...     jitter_method="replacement",
... )
\end{Code}
\end{CodeChunk}

The argument \code{jitter\_dist} accepts any \tfpshort{} distribution instance, which is used to generate pseudo-random values, and \code{jitter\_method} controls how these values are used to perturb the initial values. The current options are \code{"additive"}, \code{"multiplicative"}, and \code{"replacement"}.

\subsubsection{Debugging}

\goose{} can be configured to store additional information about the sampling process that can be helpful for debugging purposes. For example, kernel states can be recorded by passing \code{store\_kernel\_states=True} to \code{gs.LieselMCMC.run\_for\_epochs()}, and subsequently accessed with \code{get\_warmup\_kernel\_states()} on \code{gs.SamplingResults}.
For kernels that use step size tuning via dual averaging, this allows the inspection of step size convergence. For Hamiltonian Monte Carlo kernels, it further allows the inspection of the mass matrix/vector tuning process. The exact information available depends on the used kernels and the information stored in their states.

Additionally, the value of any variable (or computational node) in a \liesel{} model can be tracked and included in the samples recorded by \goose{} by passing a list of variable names in the argument \code{positions\_included} to \code{gs.LieselMCMC.run\_for\_epochs()}.

\section[Liesel-GAM]{\lieselgam{}}
\label{sec:liesel-gam}

\subsection{Introduction} \label{sec:gam-intro}

\lieselgam{} is a \proglang{Python} library that provides functionality for setting up  generalized additive model components within the \liesel{} framework, available as an extension to maintain the model-agnostic nature of \lieselmodel{} and \goose{}. The~library offers high-level abstractions that make it convenient to construct complex additive models using \liesel{}'s probabilistic modeling capabilities and to sample from their posterior using \goose{}'s MCMC infrastructure. Like the core framework, \lieselgam{} is hosted and developed on GitHub.\footnote{\url{https://github.com/liesel-devs/liesel_gam}} Documentation is available at \url{https://liesel-gam.readthedocs.io}. We import \lieselgam{} as:

\begin{Code}
>>> import liesel_gam as gam
\end{Code}

Note that, as briefly mentioned in the introduction, we use the term ``generalized additive models'' (GAMs) in a broad sense: the provided building blocks can be used not only for the mean as in traditional GAMs, but for any model parameter, enabling the specification of  generalized additive models for location, scale, and shape (GAMLSS) and developments beyond this model class. In the statistical literature, our notion of GAMs is  also referred to as structured additive distributional regression or as semi-parametric regression.

\subsection{Fundamental concepts} \label{sec:gam-fundamentals}

\subsubsection{Generalized additive distributional regression models}

In a stylized fashion, the hierarchical specification of generalized additive distributional regression models, as implemented in \lieselgam{}, is given by:
\begin{eqnarray*}
  y \mid \nuvec &\sim& \mathcal{D}(\theta_1(\nuvec), \dots,\theta_P(\nuvec)),  \\[0.5em]
  \theta_p(\nuvec) &=& h_p(\eta_p(\nuvec)), \\
  \eta_p(\nuvec) &=& f_{p1}(\nuvec; \betavec_{p1}) + \dots + f_{pL_p}(\nuvec; \betavec_{pL_p}),\\[0.5em]
  f_{pl}(\nuvec; \betavec_{pl}) &=& \mathbf{b}_{pl}(\nuvec)^\top \betavec_{pl} \\
  \betavec_{pl} \mid \tau_{pl}^2 &\sim& \mathcal{N}(0, \tau_{pl}^2\Kmat_{pl}^{-}),  \\
  \tau_{pl}^2 &\sim& \mathcal{IG}(a_{pl}, b_{pl}),
\end{eqnarray*}
where $p=1,\ldots,P$ and $l=1,\ldots,L_p$;
see \citet{Kneib2019-ModularRegressionLego} for more details.
On the first level, the parametric distribution $\mathcal{D}$ is assumed for the response variable $y$, which depends on covariates $\nuvec$ through the $P$-dimensional, covariate-dependent parameter vector $\thetavec(\nuvec)$. A wide variety of response distributions and parameters is supported by the model class. Each individual parameter $\theta_p(\nuvec)$ is related to a semi-parametric regression predictor $\eta_p(\nuvec)$ via a known response function $h_p$, which maps the predictor to the potentially constrained parameter space. The predictors are additively composed of parametric and non-parametric covariate effect terms $f_{pl}(\nuvec; \betavec_{pl})$, which are constructed as basis expansions with vectors of basis evaluations $\mathbf{b}_{pl}(\nuvec)$ and regression coefficients $\betavec_{pl}$. This framework can represent linear, spline-based, spatial, and random effects, among others. To regularize the estimation of the potentially high-dimensional regression coefficients, multivariate normal priors are assumed for $\betavec_{pl}$, where the positive semi-definite penalty matrices $\Kmat_{pl}$ are chosen to reflect the desired regularization purpose (smoothness, shrinkage, etc.). The penalty matrices $\Kmat_{pl}$ may be rank-deficient, in which case the multivariate normal prior is singular, and $\Kmat_{pl}^{-}$ denotes a generalized inverse of $\Kmat_{pl}$. The variance parameters $\tau_{pl}^2$ determine the importance of the prior (or equivalently, the amount of regularization), and are themselves assigned a hyperprior, e.g.~an inverse gamma prior with parameters $a>0$, $b>0$ \citep[see][ for a discussion of other possible hyperprior structures]{Klein2016Priors}. While \lieselgam{} uses inverse gamma priors as the default,  $\tau_{pl}^2$ can easily be specified with other priors, as we show below.

In the terminology of a \lieselmodel{} graph, the response and covariates make up the observed variables, the parameters $\theta_p(\nuvec)$, predictors $\eta_p(\nuvec)$, and bases $\mathbf{b}_{pl}(\nuvec)$ are weak variables, and the regression coefficients $\betavec_{pl}$ and variance parameters $\tau^2_{pl}$ are the model parameters.

\subsubsection[The AdditivePredictor class]{The \code{AdditivePredictor} class}

For quickly building an additive predictor $\eta_p(\nuvec)$ with multiple terms, \lieselgam{} additionally provides the \code{gam.AdditivePredictor} class, a specialized \code{lsl.Var} that allows users to add any \code{lsl.Var} object using \proglang{Python}'s \code{+=} operator and computes the sum of its inputs' values:

\begin{CodeChunk}
\begin{Code}
>>> ap1 = gam.AdditivePredictor(name="ap")
>>> ap1.value
Array(0., dtype=float32, weak_type=True)

>>> ap1 += lsl.Var.new_param(3.0, name="term1")
>>> ap1 += lsl.Var.new_param(1.0, name="term2")
>>> ap1.value
Array(4., dtype=float32, weak_type=True)
\end{Code}
\end{CodeChunk}

The \code{gam.AdditivePredictor} class is initialized with an intercept term that uses a constant prior, but any \code{lsl.Var} (or \code{None}, if no intercept should be included) can be passed to the \code{intercept} argument.

\subsubsection[The TermBuilder class]{The \code{TermBuilder} class}

\lieselgam{} makes use of the modular hierarchical structure of generalized additive models by providing  the \code{gam.TermBuilder} class for quickly setting up specialized \code{lsl.Var} objects that represent terms $f_{pl}(\nuvec; \betavec_{pl})$, including all of their parent nodes. Table~\ref{tab:gam-terms} shows a selection of the available term constructors. Table~\ref{tab:gam-terms} additionally shows common formula strings accepted by the linear constructors. Design matrices are created from the formulas using the \proglang{Python} library \pkg{Formulaic} \citep{Wardrop2026-Formulaic}.
The \code{gam.TermBuilder} populates the \code{inference} fields of the variables representing $\boldsymbol{\beta}_{pl}$ with \code{MCMCSpec} objects defining appropriate IWLS kernels.
For terms that require variance parameters, the \code{gam.TermBuilder} sets up $\tau_{pl}^2~\sim~\mathcal{IG}(a_{pl}, b_{pl})$ parameters with concentration\footnote{The concentration parameter is often also called shape, we use the \tfpshort{} terminology for the names here.} $a_{pl}=1$ and scale $b_{pl}=0.005$, following the recommendation given by \citet{Lang2004}, since, as $b_{pl} \to 0$, it approaches a constant prior on the level of $\tau_{pl}^{-2}$. \lieselgam{} sets up Gibbs samplers, updating each $\tau_{pl}^2$ from its full conditional distribution
$$
\tau_{pl}^2 \mid \cdot \sim \mathcal{IG}\bigl(a_{pl} + \nicefrac{1}{2}\operatorname{rk}(\Kmat_{pl}),\ b_{pl} + \nicefrac{1}{2} \betavec_{pl}^\top \Kmat_{pl} \betavec_{pl}\bigr),
$$
where $\operatorname{rk}(\Kmat_{pl})$ denotes the rank of $\Kmat_{pl}$. Using the collembola data from Section~\ref{sec:example} as an example, a P-spline term can be initialized with

\begin{CodeChunk}
\begin{Code}
>>> tb = gam.TermBuilder.from_df(collembola)
>>> s_lon = tb.ps("lon", k=10)
\end{Code}
\end{CodeChunk}

\begin{figure}[tb]
    \centering
    \begin{minipage}{.29\linewidth}
    \footnotesize
    \centering
    \vspace{-0.5cm}
    \code{tb.ps("lon", k=10)}
    \includegraphics[width=\linewidth]{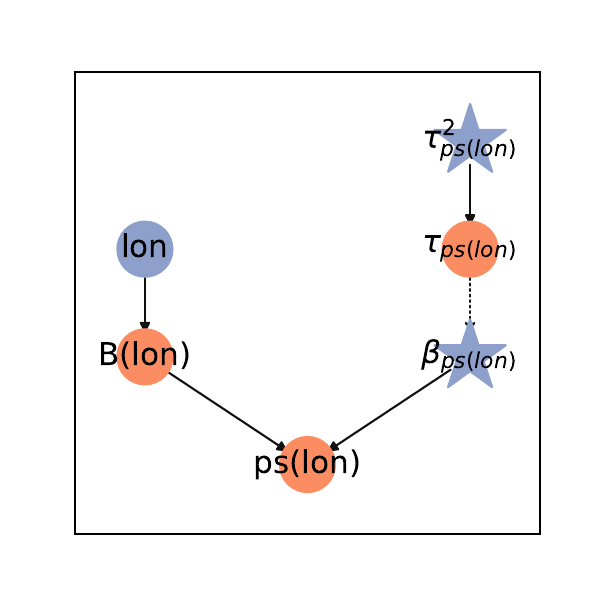}\\
    \vspace{-0.3cm}
    \includegraphics[width=0.65\linewidth]{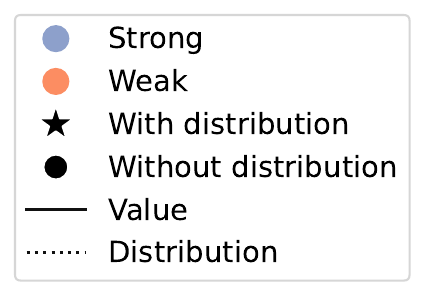}
    \end{minipage}
    \begin{minipage}{.34\linewidth}
    \footnotesize
    \centering
    \code{tb.tx(ps_lon, ps_lat)}
    \includegraphics[width=\linewidth, trim={3em 3em 3em 3em}, clip]{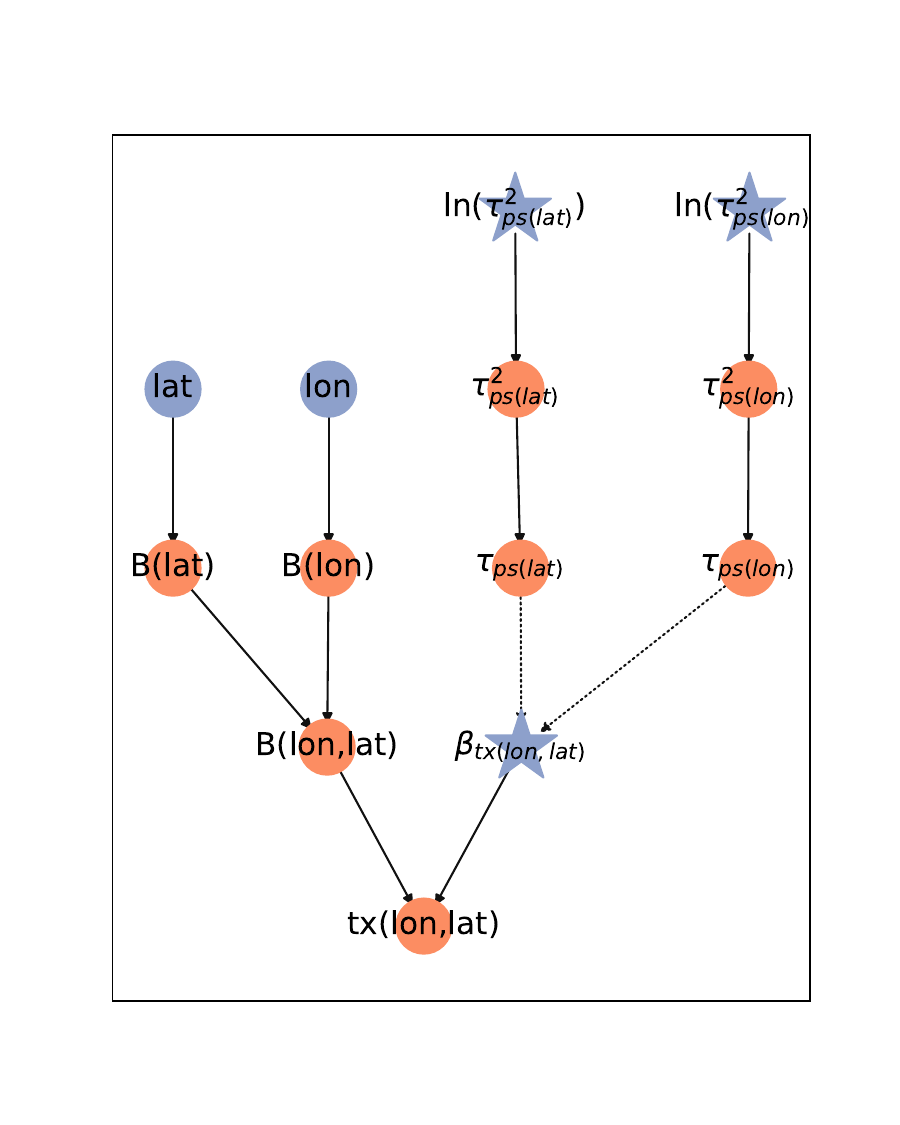}
    \end{minipage}
    \begin{minipage}{.34\linewidth}
    \footnotesize
    \centering
    \code{tb.tf(ps_lon, ps_lat)}
    \includegraphics[width=\linewidth, trim={3em 3em 3em 3em}, clip]{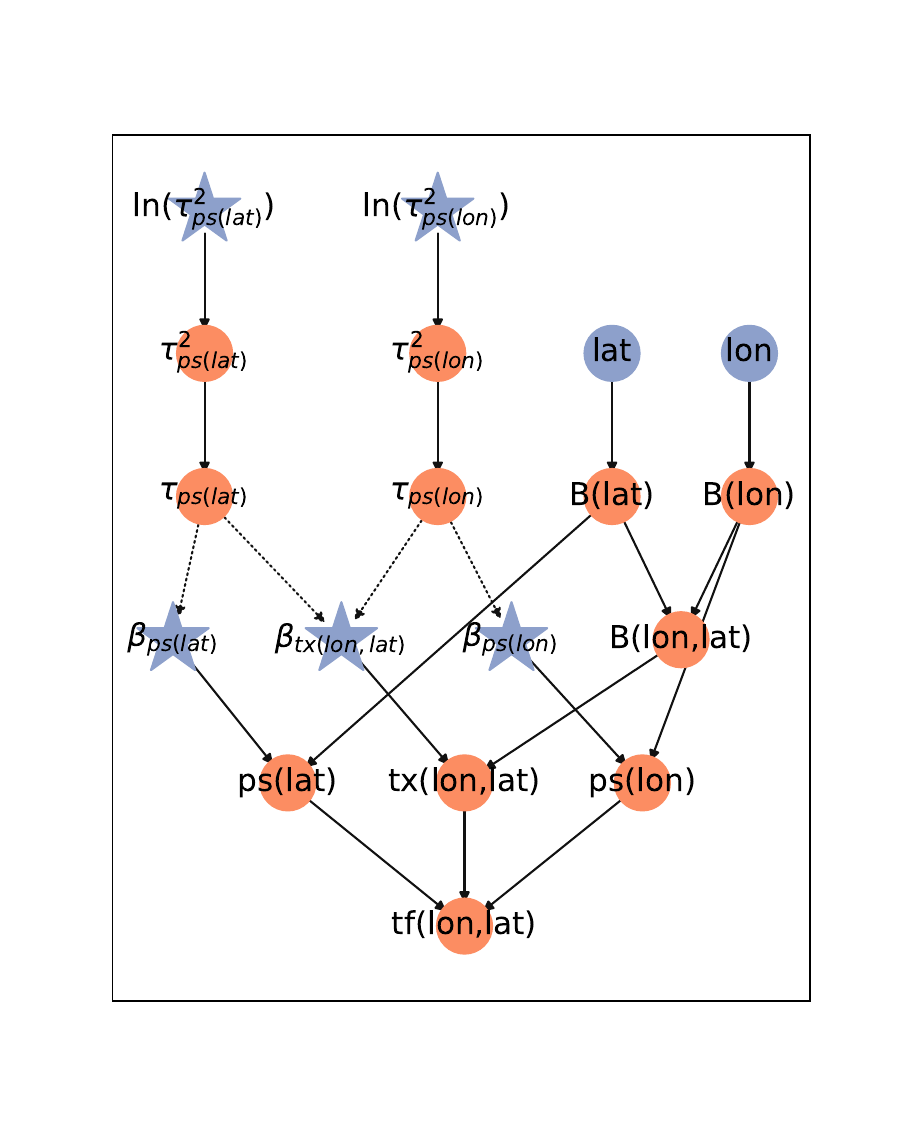}
    \end{minipage}
    \caption{Model graphs for a P-spline (left), a tensor product interaction term (middle) and a full tensor product term including main effects (right). The scale variables are taken from the marginal terms passed to the constructor. All three graphs are created using the default \code{scale} in the marginal terms.}
    \label{fig:gam-txgraph}
\end{figure}

The graph of this term can be plotted with \code{s_lon.plot()}, see Figure~\ref{fig:gam-txgraph}.
Both the default inference specification and the default smoothing scale parameter $\tau_{pl}$ can be specified on the level of the \code{TermBuilder} to be applied to all terms created with this builder.
The \code{scale} argument accepts any \code{lsl.Var}. For example, in the following code chunk we use a half-normal prior on $\tau_{pl}$, apply a \code{tfb.Exp()} bijector, and assign a NUTS kernel to sample $\log(\tau_{pl})$:
\begin{CodeChunk}
\begin{Code}
>>> scale_lat = lsl.Var.new_param(
...     1.0,
...     dist=lsl.Dist(tfd.HalfNormal, scale=10.0),
...     bijector=tfb.Exp(),
...     inference=gs.MCMCSpec(gs.NUTSKernel),
...     name="{x}",
... )
>>> s_lat = tb.ps("lat", k=10, scale=scale_lat)
\end{Code}
\end{CodeChunk}

\begin{figure}[tb]
    \centering
    \begin{minipage}{.43\linewidth}
        \includegraphics[width=\linewidth]{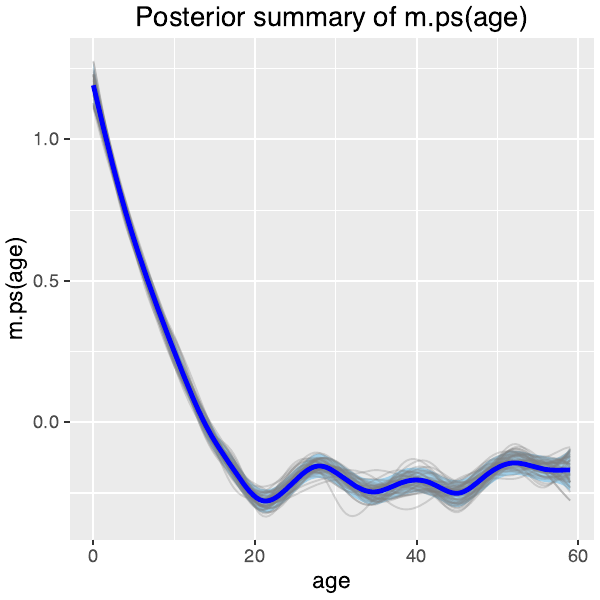}
    \end{minipage}
    \begin{minipage}{.54\linewidth}
        \includegraphics[width=\linewidth]{out/s7_mrf.pdf}
    \end{minipage}
    \caption{Examples for posterior summary plots created by \lieselgam{} with default settings. The plots use data on childhood malnutrition in Zambia between 1991 and 2001, modeled in case study II, see Section~\ref{sec:case-study-2}. The left plot was created using \code{gam.plot\_1d\_smooth()} and shows the posterior mean of a P-spline (blue line), a 90\% posterior credible band (shaded area), and 50 random posterior samples of the fitted function. The right plot was created using \code{gam.plot\_regions()} and shows the posterior mean in each district. The legend entry ``observed'' indicates that districts with observations in the data are drawn with a gray border, which are all districts in this case. Districts without observations, if present in the data, would be drawn with a red border.}
    \label{fig:case-study-2-effectplots}
\end{figure}

\subsubsection{Plots and summaries}

\lieselgam{} includes functions for quick plots and summary tables for most effect types, which are listed below. Example plots using the default settings are included in Figure~\ref{fig:a1-results} and Figure~\ref{fig:case-study-2-effectplots}. The plotting functions internally rely on the corresponding summary functions, which can be used independently. For the list below, let \code{tb = gam.TermBuilder.from\_df(dataframe)}.

\begin{itemize}
    \item \code{gam.plot\_1d\_smooth()}: One-dimensional smooths, such as \code{tb.ps}, \code{tb.np}, \code{tb.cp}. Relies on \code{gam.summarise\_1d\_smooth()}.
    \item \code{gam.plot\_2d\_smooth()}: Two-dimensional smooths, such as \code{tb.tx} and \code{tb.tf}. Relies on the function \code{gam.summarise\_nd\_smooth()}.
    \item \code{gam.plot\_1d\_smooth\_clustered()}: Clustered one-dimensional smooth terms, such as \code{tb.rs}. Relies on \code{gam.summarise\_1d\_smooth\_clustered()}.
    \item \code{gam.plot\_forest()}: Linear effects, random intercepts, and Markov random fields. Relies on cluster and linear summary helpers.
    \item \code{gam.plot\_regions()}: Markov random fields. Relies on the region summary helper.
\end{itemize}

\begin{table}[bt]
\centering
\footnotesize
\begin{tabularx}{\linewidth}{lX}
\toprule
Constructor & Description \\
\midrule
\multicolumn{2}{l}{\code{tb = gam.TermBuilder.from\_df(dataframe)}} \\
\addlinespace
\midrule
\multicolumn{2}{l}{Return variables of type \code{gam.StrctTerm} and its subtypes} \\
\addlinespace
\code{tb.slin(formula)} & Linear and categorical terms, using an i.i.d. Gaussian (ridge) prior.\\
\code{tb.ri(cluster)} & Random intercept.\\
\code{tb.ps(x)} & Smooth nonlinear functions using P-splines.\\
\code{tb.np(x)} & Purely nonlinear P-splines, i.e., without linear trend.\\
\code{tb.cp(x)} & Smooth cyclic P-splines.\\
\code{tb.kriging(x)} & Smooth spatial effects based on Gaussian processes.\\
\code{tb.mrf(x)} &  Discrete spatial effect via Markov random fields.\\
\code{tb.f()} & Accepts a custom basis function and penalty matrix. \\
\addlinespace
\midrule
\multicolumn{2}{l}{Return variables of type \code{gam.StrctInteractionTerm} and \code{gam.StrctTensorProdTerm}} \\
\addlinespace
\code{tb.tx(*marginals)} &  Interactions between multiple marginal \code{gam.StrctTerm}s. \\
\code{tb.tf(*marginals)} &  Interactions between multiple marginal \code{gam.StrctTerm}s, including lower-order terms. \\
\addlinespace
\midrule
\multicolumn{2}{l}{Return variables of other types} \\
\addlinespace
\code{tb.lin(formula, prior)} & Linear and categorical terms, defaulting to a constant prior.\\
\code{tb.rs(x, cluster)} & Random slopes for hierarchical models, corresponds to \code{x * ri(cluster)}.\\
\code{tb.vc(x, by)} &  General varying coefficient effect of \code{x}, corresponds to \code{x * by}.\\
\bottomrule
\addlinespace
Formula string &  Description \\
\midrule
\code{"a + b"} & Linear effects of \code{a} and \code{b}\\
\code{"a + b + a:b"} & Linear effects of \code{a} and \code{b} plus linear interaction.\\
\code{"a * b"} & Equivalent to \code{a + b + a:b}\\
\code{"(a + b)**n"} & Interaction of order \code{n}\\
\code{"a / b"} & Nested effect, equivalent to \code{a + a:b}\\
\code{"C(a, contr.treatment)"} & Categorical effect of \code{a} in treatment coding (default). Other codings are available, for example \code{contr.sum}, \code{contr.helmert}, \code{contr.diff}, and \code{contr.poly}.   \\
\code{"scale(a)"} & Linear effect of a, after shifting it to mean zero and scaling it to variance 1. \\
\code{"lag(a)"} & Linear effect of lagged a. \\
\bottomrule
\end{tabularx}
\caption{Selection of common \code{TermBuilder} constructors and selection of common formula strings accepted in \code{gam.TermBuilder.lin()} and \code{gam.TermBuilder.slin()} via \pkg{Formulaic}, assuming that \code{a} and \code{b} are columns in the dataframe wrapped by the used \code{gam.TermBuilder}.\label{tab:gam-terms}}
\end{table}

\subsubsection{Tensor products}

\lieselgam{} supports both isotropic and anisotropic multidimensional tensor products of additive model terms as described by \citet{Kneib2019-ModularRegressionLego} and \citet{Bach2025-AnisotropicMultidimensionalSmoothing}. The implementation relies on a custom distribution class \code{gam.MultivariateNormalStructured}, which uses the computationally efficient log density evaluation of the tensor product prior developed by \citet{Bach2025-AnisotropicMultidimensionalSmoothing}. Tensor product terms are specified by combining marginal terms in \code{gam.TermBuilder.tx()} as
\begin{CodeChunk}
\begin{Code}
>>> tb.tx(s_lon, s_lat)
\end{Code}
\end{CodeChunk}
for pure interactions and \code{gam.TermBuilder.tf()} for full tensor products including main effects and lower-order interactions. Figure~\ref{fig:gam-txgraph} shows the term graphs for both variants using the longitude and latitude P-spline terms initialized above. The tensor product constructors accept an arbitrary number of arguments. As marginals, terms of type \code{gam.StrctTerm} and its subtypes are allowed, see Table~\ref{tab:gam-terms}.

\subsection{Advanced topics} \label{sec:gam-advanced}

\subsubsection{Linear constraints}

\code{gam.StrctTerm} allows users to apply general linear constraints via reparameterization with the method \code{gam.StrctTerm.constrain()}, which accepts a constraint matrix $\mathbf{A}$ and reparameterizes the basis function and the prior for $\boldsymbol{\beta}_{pl}$ to enforce the constraint $\mathbf{A}\boldsymbol{\beta}_{pl} = \mathbf{0}$ \citep[see][ for details]{Kneib2019-ModularRegressionLego}. This feature makes it especially easy for users to either apply custom linear constraints to the builtin effect types, or to apply constraints to their own custom basis implementations. By default, the constructors available via \code{gam.TermBuilder} automatically apply suitable constraints (usually sum-to-zero constraints). This default behavior can be turned off by supplying the argument \code{absorb\_cons=False} to terms that accept it.

\subsubsection{Other reparameterizations}

For most terms, \lieselgam{} defaults to using penalty matrices that are diagonalized via eigenvalue decomposition and scaled using their infinity norm.  While these settings can lead to more numerical stability and more favorable posterior geometry, they may be turned off using the arguments \code{diagonalize\_penalty} and \code{scale\_penalty} in the \code{gam.TermBuilder} constructors. Diagonalization and scaling may also be applied to custom bases by the corresponding methods on \code{gam.StrctTerm}.

The argument and corresponding method \code{factor_scale} can be used to reparameterize the implementation of the coefficient prior of terms that use an identity matrix for $\Kmat_{pl}$ into the form
\begin{align} \label{eq:reparam-factor_scale}
\boldsymbol{\beta}_{pl}  = \tau_{pl} \tilde{\boldsymbol{\beta}}_{pl}, \qquad
\tilde{\boldsymbol{\beta}}_{pl} \sim \mathcal{N}(\mathbf{0}, \Kmat_{pl}^{-}).
\end{align}
This reparameterization, sometimes called non-centered parameterization, in which $\tilde{\betavec}_{pl}$ can be understood as a standardized coefficient, can be helpful for example for sampling high-dimensional random intercepts using NUTS \citep[see, for example,][, Chapter 27.7]{StanUserGuide2-38}.

\subsubsection[Interfacing with mgcv and JIT-compatibility]{Interfacing with \pkg{mgcv} and JIT-compatibility}

For many of the supported term types, \lieselgam{} uses \pkg{ryp} \citep{WainbergRyp} for interfacing internally with the \proglang{R} package \pkg{mgcv} \citep{Wood2022} to extract basis functions and penalty matrices, converting them to \pkg{NumPy} arrays.
For users, this integration works seamlessly in the background, allowing \lieselgam{} to leverage the mature and sophisticated setup functionality of \pkg{mgcv} within a fully \proglang{Python}-based workflow on the user side. The methods that interface with \proglang{R} are, however, not compatible with \jax{}'s just-in-time compilation and automatic differentiation. In most common use cases, this incompatibility poses no problems as the inputs to basis functions typically remain constant throughout MCMC sampling in generalized additive models. If users wish to use bases with dynamic inputs, they generally need to provide a \jax{}-compatible pure \proglang{Python} implementation for use in \code{gam.TermBuilder.f()}. The same limitation applies to the linear bases provided via \pkg{Formulaic} in \code{gam.TermBuilder.lin()}.

\section[Case study I]{Case study I: Collembola occupancy model}
\label{sec:case-study-1}

We now expand the logistic collembola detection model presented in Section~\ref{sec:example} into an occupancy model that incorporates uncertainty about species detection and includes two indices of species-richness. Species richness is an indicator of biodiversity that can be evaluated on different geographical scales, giving rise to $\alpha$-richness (locally) and $\gamma$-richness (on a landscape level). This example shows a straightforward handling of a partially unobserved variable of interest, the implementation of a custom Gibbs sampler for this discrete variable, and the inclusion of derived measures directly into the model.

\subsection{Modeling occupancy}

If $\mathtt{detection}_{ij}=0$, this may mean that species $j$ does not occupy plot $i$, or that species $j$ does occupy plot $i$ but no individuals of this species were found in the collected soil sample. To account for this uncertainty in the model, we introduce a partially unobserved binary variable $\mathtt{occupancy}_{ij} \in \{0, 1\}$, where $\mathtt{occupancy}_{ij} = 1$ if species $j = 1, \dots, M$ occupies plot $i = 1, \dots, N$. Positive detection ($\mathtt{detection}_{ij}=1$) implies occupancy, but negative detection does not necessarily imply \textit{no occupancy}, so occupancy is unobserved in this case. The updated response model is
\begin{align}
\mathtt{occupancy}_{ij} &\sim \operatorname{Bernoulli}(\psi_{ij}) \\
\mathtt{detection}_{ij} & \sim \operatorname{Bernoulli}(\mathtt{occupancy}_{ij} \cdot \pi),
\end{align}
where $\pi$ is the probability of detecting a species if it occupies a plot and $\psi_{ij}$ is modeled on the logit level as in Section~\ref{sec:example}. For demonstration purposes, we assume a fixed $\pi=0.9$ for this example.\footnote{
The detection probability is not identifiable from this dataset alone. The number $0.9$ is an ad-hoc choice for demonstration purposes based on the following reasoning. For each plot, only a single soil core with a diameter of 5 cm was collected, so detection may not be perfect. Still, Collembola are very abundant in temperate coniferous forests \citep{Junggebauer2025-TemporalDynamicsStability}, so the true detection probability is likely to be close to $1$. For a conclusive analysis, a better estimate for the detection probability and its uncertainty should be developed.
} During MCMC sampling, we draw pseudo-observations for the unobserved occupancy indicators using a Gibbs sampler, where
\begin{equation} \label{eq:cs1-fullconditional}
P(\mathtt{occ}_{ij} = 1 \mid \boldsymbol{\theta}_{-ij}, \boldsymbol{y}) = \frac{p(\boldsymbol{\theta}_{-ij}, \mathtt{occ}_{ij} = 1 \mid \boldsymbol{y})}{
p(\boldsymbol{\theta}_{-ij}, \mathtt{occ}_{ij} = 1 \mid \boldsymbol{y})
+
p(\boldsymbol{\theta}_{-ij}, \mathtt{occ}_{ij} = 0 \mid \boldsymbol{y})
},
\end{equation}
if $\mathtt{detection}_{ij}=0$ and $1$ otherwise, $\boldsymbol{y}$ denotes the values of all observed variables in the model, and we use $\boldsymbol{\theta}_{-ij}$ to generically represent all model parameters except the focal $\mathtt{occ}_{ij} = \mathtt{occupancy}_{ij}$.

The model graph is shown in Figure~\ref{fig:cs1-results} (left).
In the implemented \liesel{} model, we split the occupancy indicators into an observed and an unobserved subset and use a \code{gs.GibbsKernel} to carry out sampling for the unobserved occupancy indicators. The split can be achieved with simple indexing operations wrapped by \code{lsl.Var.new\_calc} variables; for example, for the unobserved indicators, we use
\begin{CodeChunk}
\begin{Code}
>>> unobs_index = jnp.flatnonzero(~jnp.asarray(collembola["presence"]))
>>> logit_psi_unobserved = lsl.Var.new_calc(
...     lambda x: x[unobs_index], x=logit_psi, name=r"$logit(\psi)_{unobs}$"
... )
>>> unobserved_occupancy = lsl.Var.new_param(
...     collembola["presence"].astype(int)[unobs_index],
...     dist=lsl.Dist(tfd.Bernoulli, logits=logit_psi_unobserved),
...     name="unobs_occupancy",
... )
\end{Code}
\end{CodeChunk}

The Gibbs kernel is then implemented by writing a transition function that loops over all unobserved occupancy indicators to draw values from their full conditionals (\ref{eq:cs1-fullconditional}), and attaching a suitable constructor function to \code{unobserved\_occupancy.inference}. The code is included in the supplementary material.

\subsection{Including species richness indicators}

The concept of $\alpha$-richness is a local measure, computed as the number of species found per plot. In contrast, $\gamma$-richness captures larger geographic areas, so it is computed as the number of species found in northern or southern Lower Saxony. To account for uncertain occupancy in the absence of detection, the richness measures are computed based on the partially estimated occupancy indicators. For example, $\alpha$-richness is initialized as
\begin{CodeChunk}
\begin{Code}
>>> alpha_richness = lsl.Var.new_calc(
...     lambda x: jnp.bincount(plot_indicators, weights=x, length=n_plots),
...     x=occupancy,
...     name=r"$\alpha$-richness",
... )
\end{Code}
\end{CodeChunk}

The object \code{plot_indicators} identifies which row of the data belongs to which plot.
Including the measures in the \code{lsl.Model} via
\begin{CodeChunk}
\begin{Code}
>>> model = lsl.Model(detection, alpha_richness, gamma_richness)
\end{Code}
\end{CodeChunk}
greatly simplifies posterior inference for these measures, because their posterior distributions can be assessed by propagating samples from the posterior distributions of the model parameters through the model graph via \code{lsl.Model.predict}.

\subsection{Results}

\begin{figure}[t]
    \centering
    \begin{minipage}{.32\linewidth}
        \includegraphics[width=\linewidth]{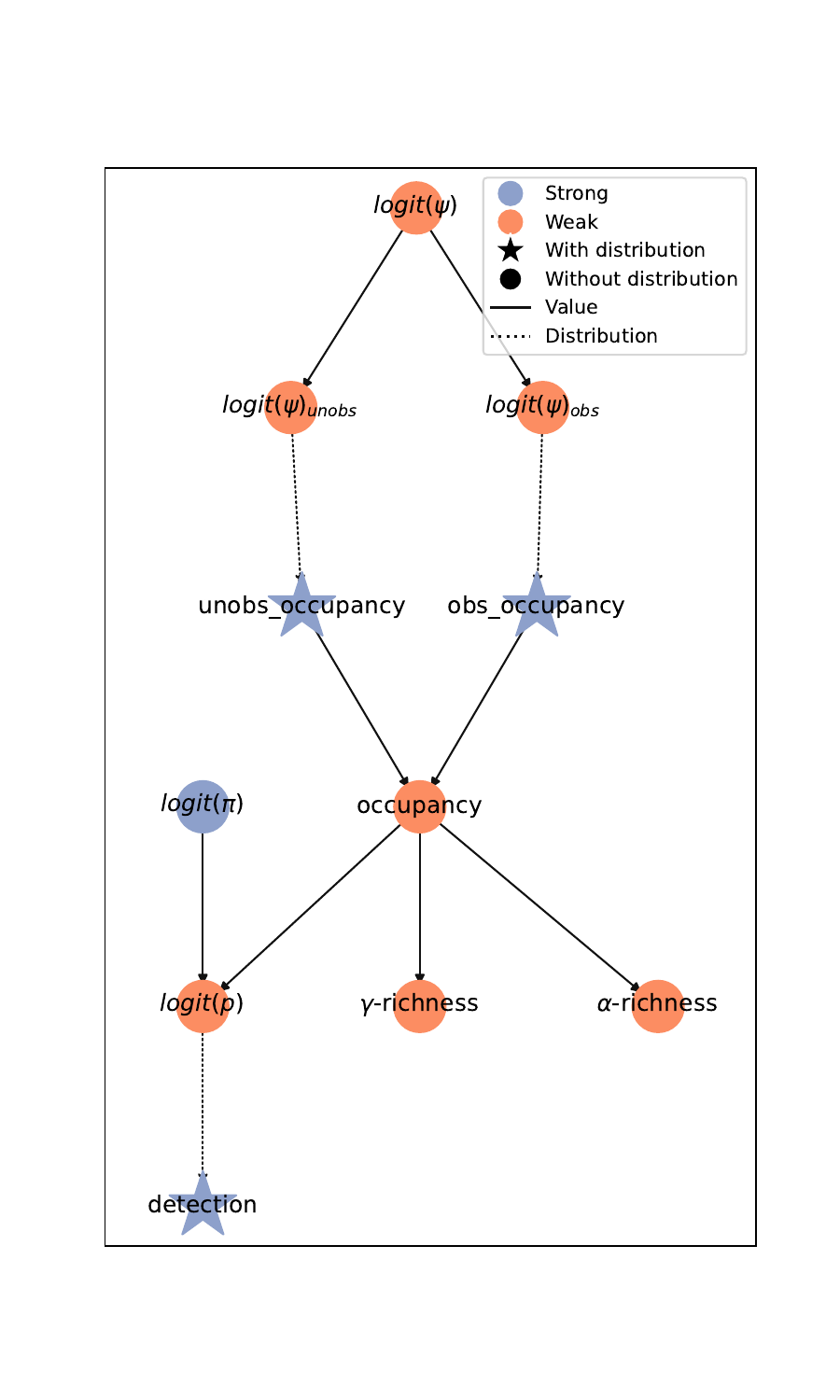}
        \vspace{0.07cm}
    \end{minipage}
    \begin{minipage}{.32\linewidth}
        \includegraphics[width=\linewidth]{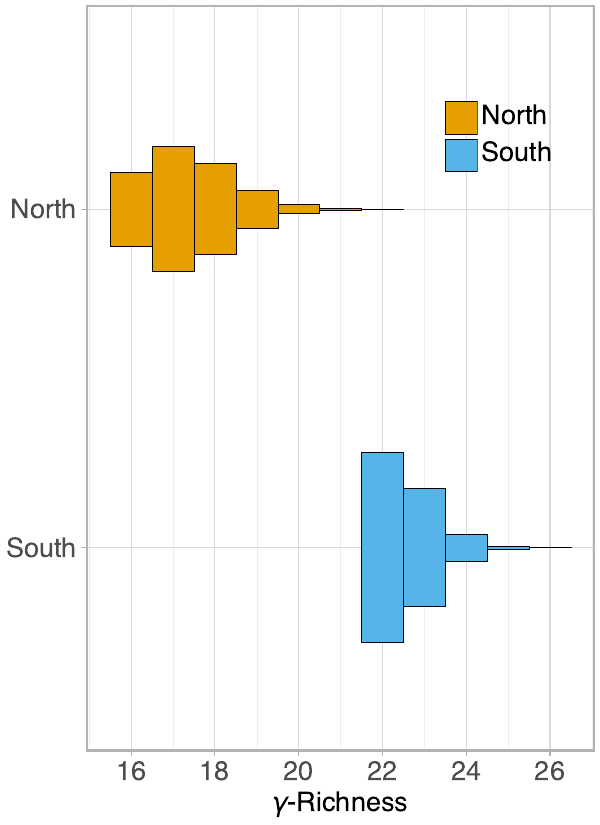}
    \end{minipage}
    \begin{minipage}{.32\linewidth}
        \includegraphics[width=\linewidth]{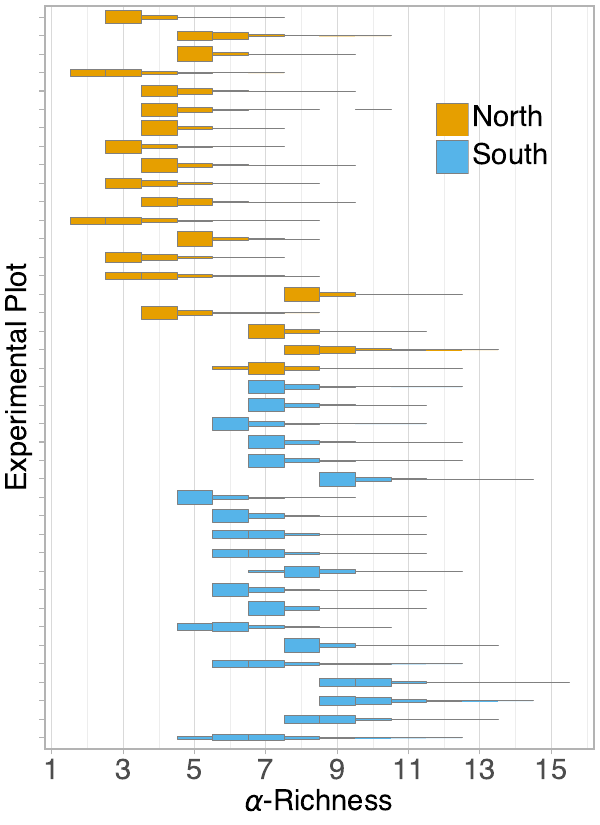}
    \end{minipage}
    \caption{Left: Response model of the collembola occupancy model in case study I. For numerical stability, probabilities are handled on logit level in the model. Middle: Posterior distribution of $\gamma$-richness, indicating species diversity by landscape. Right: Posterior distribution of $\alpha$-richness, indicating species diversity by experimental plot. Both measures show a consistently greater diversity in southern Lower Saxony. In the middle and right panel, the heights of the tiles are given by the relative posterior frequencies of the values on the x-axes.}
    \label{fig:cs1-results}
\end{figure}

The posterior distributions for $\gamma$- and $\alpha$-richness are displayed in the middle and right panels of Figure~\ref{fig:cs1-results}, respectively. Both on the highly aggregated $\gamma$-level, and the local $\alpha$-level, we can observe the species richness is estimated to be consistently higher in southern Lower Saxony than in the north. The linear effects for the area potentially available for Norway spruce (posterior mean $\hat{\beta} \approx 0.09$, 90\% credible interval $[-0.55; 0.71]$) and Douglas fir (posterior mean $\hat{\beta} \approx -0.04$, 90\% credible interval $[-0.70; 0.67]$) do not indicate a clearly detectable meaningful influence of forest composition on collembola occupancy in this model.

\section[Case study II]{Case study II: Comparing sampling strategies}
\label{sec:case-study-2}

In this case study, we make use of \goose{}'s flexibility to explore the performance of different MCMC sampling strategies using
a dataset on childhood malnutrition in Zambia, collected as part of representative household surveys conducted in developing countries. The data is freely available at \url{https://dhsprogram.com/}. We use a total of $N=14\,151$ observations collected in 1992, 1996, and 2001, and consisting of the variables $\mathtt{z}_i$ (a measure of chronic malnutrition, lower values indicate  more severe malnutrition; we use a standardized version with mean zero and variance 1), $\mathtt{age}_i$ (a child's age in months), $\mathtt{bmi}_i$ (the mother's body mass index), and $\mathtt{district}_i$ (the district of residence in Zambia, 55 districts total); $i = 1, \dots, N$. We set up a generalized additive distributional regression model with Gaussian response $\mathtt{z}_i \sim \mathcal{N}(\mu_i, \sigma_i^2)$ and
\begin{align}
    \mu_i & = \beta_0^{\mu} + f^{\mu}_1(\mathtt{age}_i) + f^{\mu}_2(\mathtt{bmi}_i) + f^{\mu}_{\text{spat}}(\mathtt{district}_i) \\
    \log \sigma_i & = \beta_0^{\sigma} + f^{\sigma}_1(\mathtt{age}_i) + f^{\sigma}_2(\mathtt{bmi}_i),
\end{align}
where $f_1$ and $f_2$ are modeled through P-splines \citep{Brezger2005} in both predictors and $f^{\mu}_{\text{spat}}$ is a discrete spatial effect modeled as a Gaussian Markov random field \citep{Rue2005-GaussianMarkovRandom}. We use $\tau^2 \sim \mathcal{IG}(1, 0.005)$ priors for all five smoothing variance parameters.

\subsection{Sampling strategies}

\begin{figure}[tb]
    \centering
    \includegraphics[width=\linewidth, trim={1em 9.5em 1em 7em}, clip]{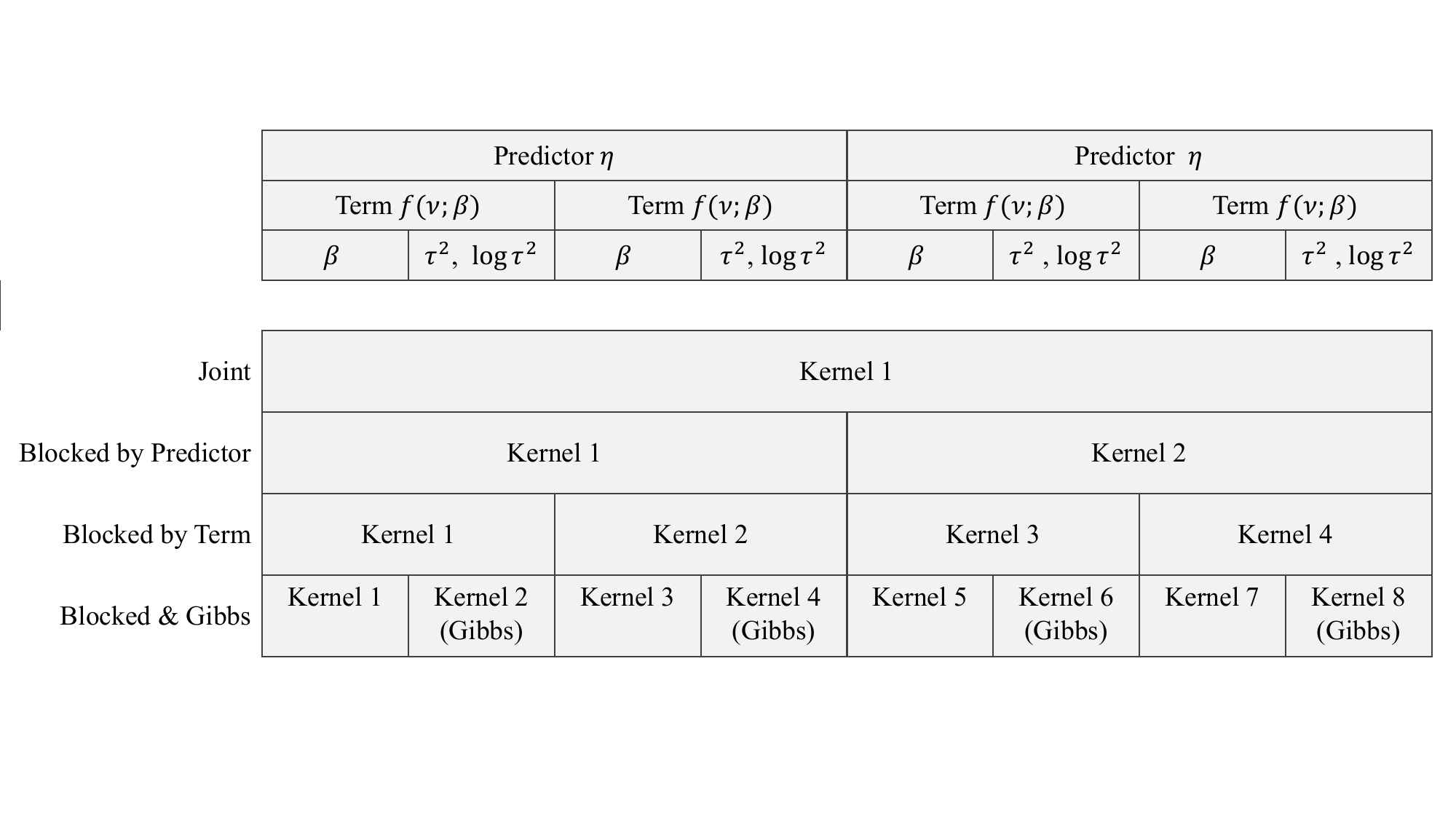}
    \caption{Conceptual overview of sampling strategies used in case study II. The generic ``Kernel'' is replaced by an IWLS, an HMC, or a NUTS kernel. All strategies except for the \textit{Blocked \& Gibbs} strategy operate on the level of $\log \tau^2$. The intercepts are treated as individual terms.}
    \label{fig:cs2-strategies}
\end{figure}

We compare four sampling strategies using three different MCMC kernels available in \goose{}, namely an Iteratively Weighted Least Squares (IWLS) kernel, a Hamiltonian Monte Carlo (HMC) kernel, and a No-U-Turn Sampler (NUTS) kernel. Figure~\ref{fig:cs2-strategies} shows an overview of the sampling strategies.

The \textit{Blocked \& Gibbs} strategy with an IWLS kernel for the $\betavec$ blocks corresponds to the default in \lieselgam{} and is a popular strategy for generalized additive models; see \citet{Klein2015-BayesianStructuredAdditiveb, Kneib2019-ModularRegressionLego,Umlauf2021}. As the IWLS kernel involves second derivatives for proposal generation, it is interesting to compare its performance to gradient-based MCMC algorithms such as HMC or NUTS that avoid second derivatives. HMC and NUTS have been popularized by software like \stan{} \citep{StanUserGuide2-38} and \pymc{} \citep{Salvatier2016}, and are known to work well in many applications \citep[, Chapter 30]{MacKay2003}.

For the IWLS kernel, we limit the consideration to the \textit{Blocked by Term} and \textit{Blocked \& Gibbs} strategies because we anticipate little benefit from taking correlation between terms into account, and the proposal distribution scales cubically in the dimension of the parameter block. For the HMC kernel, we use 50 integration steps in the leapfrog algorithm. For the NUTS kernel, we use a maximum tree depth of $10$. Both HMC and NUTS are employed with a diagonal metric, which is tuned during adaptation. In all kernels, the step size is tuned using the dual averaging algorithm during adaptation to an acceptance probability of 80\%. Different sampling strategies can be set up easily with \goose{}. For example, the joint strategy can be implemented with the following function:

\begin{CodeChunk}
\begin{Code}
>>> def strategy_joint(model: lsl.Model, kernel_constructor, **kwargs):
...     model = model.copy()
...     for k, v in model.parameters.items():
...         if "tau" in k:
...             v.biject(tfb.Exp(), inference="drop")
...     for param in model.parameters.values():
...         param.inference = gs.MCMCSpec(
...             kernel_constructor, kernel_group=1, kernel_kwargs=kwargs
...         )
...     return model
\end{Code}
\end{CodeChunk}

\subsection{Results}

We ran each sampling strategy using $2000$ adaptation and posterior iterations in four parallel chains, totaling $8000$ posterior iterations. Sampling was carried out on a 2021 14 inch MacBook Pro with an M1 chip with \liesel{} v0.5.0, \lieselgam{} v0.1.4, \jax{} v0.10.1, and \pkg{BlackJAX} v1.5. Posterior summary plots of $f^{\mu}_1(\mathtt{age}_i)$ and $f^{\mu}_{\text{spat}}(\mathtt{district}_i)$ from the \textit{IWLS: Blocked \& Gibbs} condition are included in Figure~\ref{fig:case-study-2-effectplots}.

\paragraph{Convergence.}
Three strategies showed problematic convergence diagnostics, indicated by large $\hat{R}$ for at least one parameter or parameter block: \textit{IWLS: Blocked by Term}, \textit{HMC: Joint}, and \textit{HMC: Blocked \& Gibbs}. The other strategies all converged to the same posterior distribution; albeit the two $\log(\tau^2)$ belonging to $f^{\mu}_2(\mathtt{bmi}_i)$ and $f^{\sigma}_2(\mathtt{bmi}_i)$ still showed problematic $\hat{R}$ in \textit{NUTS: Joint} at $\hat{R} \approx 1.18$ and $\hat{R} \approx 1.19$, respectively.

\begin{figure}[tb]
    \centering
    \begin{minipage}{.41\linewidth}
        \includegraphics[width=\linewidth]{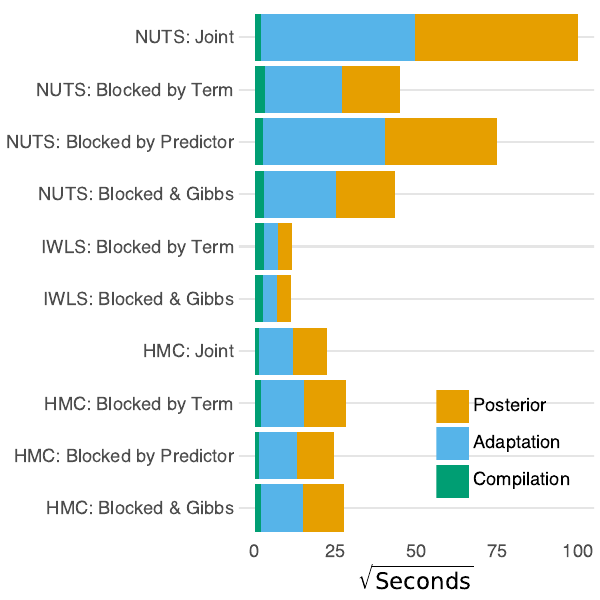}
    \end{minipage}
    \begin{minipage}{.58\linewidth}
        \includegraphics[width=\linewidth]{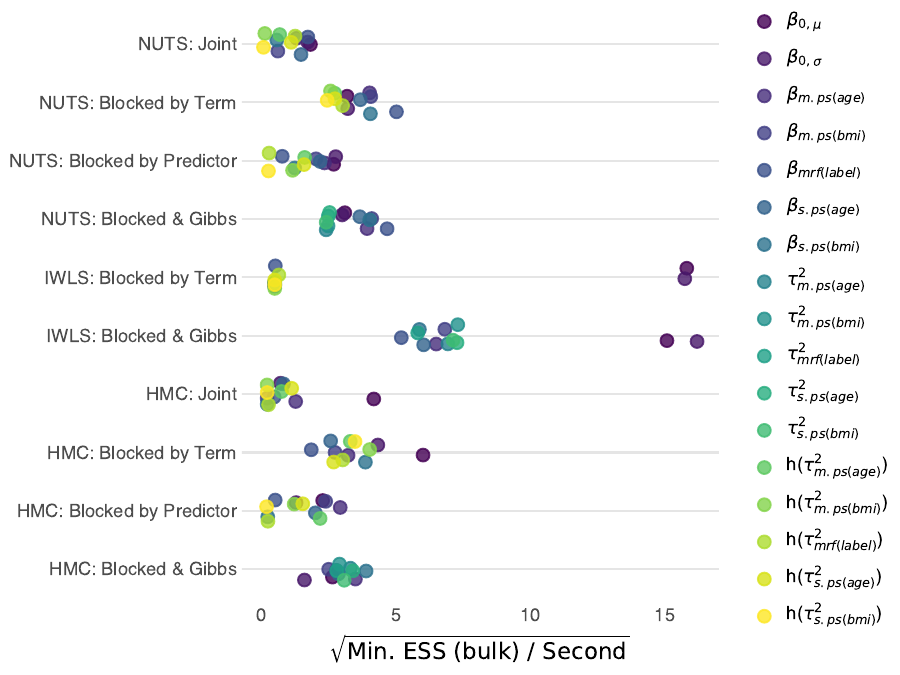}
    \end{minipage}
    \caption{Results of case study II, comparing different sampling strategies for a distributional regression model. Left: Sampling duration in seconds by phase. Right: Blockwise minimum effective samples per second (bulk), where each parameter is taken as one block. Points are placed with slight vertical random jitter, but no horizontal jitter. Both plots show the square-root of the target quantity to improve the visibility of differences.}
    \label{fig:cs2-results}
\end{figure}

\paragraph{Runtime and effective sample size.}

Figure~\ref{fig:cs2-results} shows the total runtime in seconds (left) and the effective sample size (bulk) achieved per second (right). The results indicate striking efficiency differences. The IWLS strategies both finished in $<45$ seconds, while the HMC strategies took between $208$ (\textit{Joint}) and $338$ seconds (\textit{Blocked \& Gibbs}) and the NUTS strategies took between $832$ (\textit{Blocked \& Gibbs}) and $4588$ seconds (\textit{Joint}). In the \textit{Blocked \& Gibbs} strategy, the shorter runtime of the IWLS kernel translates to superior efficiency visible in high numbers of effective samples drawn per second. The exceptionally high runtime of the NUTS strategies can be partially attributed to the algorithm reaching the maximum tree depth in many iterations, meaning that the sampler never observes the namesake ``U-turn'' during proposal generation. For example, in the \textit{Joint} strategy, NUTS reached the maximum tree depth in 95\% of posterior iterations. Such issues can be caused by difficult posterior geometries and might be solved or reduced by reparameterization, the use of a Riemann metric \citep{Girolami2011}, or different hyperparameter settings. We did not explore such options further in this case study.

\paragraph{Discussion.} The results demonstrate that the choice of sampling strategy can have a significant impact on the efficiency of MCMC sampling. They should not be interpreted to show that the \textit{IWLS: Blocked \& Gibbs} strategy is generally superior for generalized additive models, as the performance of HMC and NUTS could be caused by idiosyncrasies of the data, and they could potentially be improved by using different hyperparameters. We have indeed seen NUTS perform well for generalized additive models in other analyses, for example in case study I. The results of case study II may rather be interpreted as showing that it can be beneficial to easily have different options available. \goose{} provides such options.

\section{Conclusion}
\label{sec:conclusion}

The \proglang{Python} library \liesel{} is a probabilistic programming framework that allows users to express Bayesian models as directed acyclic graphs and to build custom MCMC algorithms using just-in-time compilation and automatic differentiation through \jax{}. Models can be probed and modified in a feature-rich API in interactive \proglang{Python} sessions. MCMC algorithms can be created by blocking and combining the efficient builtin MCMC kernels or by implementing custom MCMC kernels.
The add-on library \lieselgam{} additionally provides generalized additive model components, but we wish to emphasize that \liesel{}'s scope extends to any Bayesian model that can be expressed as a directed acyclic graph.
In recent years, it has proved its effectiveness in several studies, e.g., by \citet{marquesVariancePartitioningMultilevel2023}, \citet{Riebl2023}, \citet{Marques2025bsp}, \citet{Brachem2025-BayesianPenalizedTransformation}, \citet{Brachem2026-DataEfficientGenerativeModeling}, and \citet{Queiroz2026-BayesianStructuredAdditive}.

We are continuously working on further developing \liesel{}; for example, we are currently developing a module that allows for automatic differentiation variational inference (ADVI) on \liesel{} models, using flexible variational distributions that are themselves implemented in \liesel{}.
With its combination of features, \liesel{} offers a distinctive toolkit for statisticians to quickly iterate the development and testing of novel models and perform inference using state-of-the-art MCMC sampling techniques.
In this way, \liesel{} helps to address the central tension motivating this work by combining modeling flexibility, computational efficiency, and the availability of reusable model components within a single framework.

\section*{Acknowledgments}

HR, JB and PFVW contributed equally.
The authors gratefully acknowledge the financial support from the German Research Foundation~(DFG) under grant number 443179956. TK and HR were additionally supported by the DFG through Research Training Group (RTG) 2300. We also extend our sincere appreciation to all the contributors to the \liesel{} software framework, with a special acknowledgment of Alex Afanasev, and Joel Beck for their contributions. Additionally, we wish to thank Isa Marques for her willingness to incorporate \liesel{} into her work when the software was in its early stages and Francisco Felipe de Queiroz for extensively testing \lieselgam{}.

\bibliography{ref}

\clearpage
\begin{appendix}

\section[Details on the Var and Model classes]{Details on the \code{Var} and \code{Model} classes} \label{app:model-details}

Tables~\ref{tab:var-and-dist} and~\ref{tab:model} provide further details on the available attributes and methods of the classes \code{Var}, \code{Dist}, and \code{Model}.

\begin{table}[b]
    \centering

    \begin{tabularx}{\linewidth}{lX}
\toprule
   Name  & Description \\
   \midrule
   \textsf{\textbf{lsl.Var}} & \\
   \code{Var.name} & Variable name.\\
   \code{Var.value} & Current variable value.\\
   \code{Var.log\_prob} & Current log probability corresponding to \code{Var.value}.\\
   \code{Var.inference} & Inference specification, e.g. an instance of \code{gs.MCMCSpec}.\\
   \code{Var.value\_node} & Computational node, wrapping the variable's value.\\
   \code{Var.dist\_node} & Distribution node, a \code{lsl.Dist} object or \code{None}.\\
   \addlinespace
   \code{Var.parameter} & Boolean, indicating whether the variable is a parameter.\\
   \code{Var.observed} & Boolean, indicating whether the variable is observed.\\
   \code{Var.strong} & Boolean, indicating whether the variable is strong.\\
   \code{Var.weak} & Boolean, indicating whether the variable is weak.\\
   \addlinespace
   \code{Var.biject()} & Applies a bijective transformation to the variable and transforms its distribution accordingly.\\
   \code{Var.bijected\_var} & Reference to the bijected variable, if \code{lsl.Var.biject()} has been called.\\
   \code{Var.plot()} & Plots the model graph of this variable's sub-model, including all parent variables.\\
   \code{Var.predict()} & Computes the variable's value given a \code{samples} dictionary and, potentially a \code{newdata} dictionary.\\
   \code{Var.sample()} & Hierarchically draws samples from all random variables in this variable's sub-model.\\
\addlinespace
   \textsf{\textbf{lsl.Dist}} & \\
   \code{Dist.biject\_parameters()} & Applies bijective transformations to parameters of this distribution via \code{Var.biject()}.\\

 \bottomrule
\end{tabularx}
\caption{Overview of central \code{Var} and \code{Dist} attributes and methods.\label{tab:var-and-dist}}
\end{table}

\begin{table}[t]
    \centering
    \begin{tabularx}{\linewidth}{lX}
\toprule
   Name  & Description \\
   \midrule
   \code{Model.log\_prob} & The current joint log probability of all random variables in the model.\\
   \code{Model.log\_prior} & The current joint log probability of all parameter variables in the model.\\
   \code{Model.log\_lik} & The current joint log probability of all observed variables in the model.\\
   \code{Model.state} & The current model state.\\
   \addlinespace
   \code{Model.parameters} & A dictionary of all parameter variables in the model.\\
   \code{Model.observed} & A dictionary of all observed variables in the model.\\
   \code{Model.vars} & A dictionary of all variables in the model.\\
   \code{Model.nodes} & A dictionary of all nodes in the model's low-level computational graph.\\
   \addlinespace
   \code{Model.plot()} & Plots the model graph.\\
   \code{Model.diagnose()} & Assembles a dataframe with diagnostic information.\\
   \code{Model.predict()} & Computes the model state given a \code{samples} dictionary and, potentially a \code{newdata} dictionary.\\
   \code{Model.sample()} & Hierarchically draws samples from all random variables in this model.\\
   \addlinespace
   \code{Model.copy()} & Returns a deep copy of the model.\\
   \code{Model.add()} & Adds a variable to the model.\\
   \code{Model.replace()} & Replaces all usages of a variable in the model with another variable.\\
   \code{Model.join()} & Adds all variables from other models to this model.\\
   \addlinespace
   \code{Model.update\_state()} & Pure function for updating the model state. \\
   \code{Model.extract\_position()} & Extracts a name-value dictionary from a model state.\\

 \bottomrule
\end{tabularx}
\caption{Overview of central \code{lsl.Model} attributes and methods.\label{tab:model}}
\end{table}

\end{appendix}

\end{document}